\documentclass[journal,draftcls,onecolumn,twoside,11pt]{IEEEtran}
\normalsize
\ifCLASSINFOpdf
\else
\fi

\usepackage{amsmath}
\usepackage{cite}
\usepackage{graphicx}
\usepackage{caption}
\usepackage{subcaption}
\usepackage{float}
\usepackage{setspace}
\usepackage{siunitx}
\usepackage{multirow}
\restylefloat{figure}
\input{epsf}
\begin{document}
\doublespacing

\title{Abnormality Detection in Correlated Gaussian Molecular Nano-Networks: Design and Analysis}

\author{Siavash~Ghavami
        and~Farshad~Lahouti
\thanks{S. Ghavami is with the Department of Electrical \& Computer Engineering, University of Minnesota, USA and F. Lahouti is with the Electrical Engineering Department, California Institute of Technology, USA; E-mail: sghavami@umn.edu,lahouti@caltech.edu.}
}

\markboth{IEEE Transactions on NanoBioScience}%
{Submitted paper}

\maketitle
\vspace{-3.1\baselineskip}
\begin{abstract}
A nano abnormality detection scheme (NADS) in molecular nano-networks is studied. This is motivated by the fact that early detection of diseases such as cancer play a crucial role in their successful treatment. The proposed NADS is in fact a two-tier network of sensor nano-machines (SNMs) in the first tier and a data-gathering node (DGN) at the sink. The SNMs detect the presence of competitor cells (abnormality) by variations in input and/or parameters of a nano-communications channel (NCC). The noise of SNMs as their nature suggest is considered correlated in time and space and herein assumed additive Gaussian. In the second step, the SNMs transmit micro-scale messages over a noisy micro communications channel (MCC) to the DGN, where a decision is made upon fusing the received signals. We find an optimum design of detectors for each of the NADS tiers based on the end-to-end NADS performance. The detection performance of each SNM is analyzed by setting up a generalized likelihood ratio test. Next, taking into account the effect of the MCC, the overall performance of the NADS is analyzed in terms of probabilities of misdetection and false alarm. In addition, computationally efficient expressions to quantify the NADS performance is derived by providing respectively an approximation and an upper bound for the probabilities of misdetection and false alarm. This in turn enables formulating a design problem, where the optimized concentration of SNMs in a sample is obtained for a high probability of detection and a limited probability of false alarm. The results indicate that otherwise ignoring the spatial and temporal correlation of SNM noise in the analysis, leads to an NADS that noticeably underperforms in operations.
\end{abstract}

\vspace*{-0.25in}

\IEEEpeerreviewmaketitle

\section{Introduction}
Cancer is a leading cause of death in the world and accounts for about 13\% of all death per annum~\cite{1}. In addition, it incurs serious disability and emotional challenges to the people and heavily affects them financially~\cite{2}. There have been many advances by significant technological innovations in the treatment of cancer. However, success is still a distant goal in this direction. Indeed, research in early detection and diagnostics of cancer and their associated enabling technologies are of extraordinary importance.

It is believed that the overall behavior of cancer is determined with genes expressions and/or proteins. In fact, proteomic data and collective functions of proteins are known to directly set the cell function. Hence, modeling and analysis of genomic and proteomic data using micro array and mass spectrometry technologies have found various applications in cancer studies~\cite{3}. In~\cite{4}, an interesting review of technologies for nano-scale cancer bio-molecular detection using proteomic and genomic approaches is presented. In~\cite{5}, application of nano-technologies for bio-molecular detection and medical diagnostics is studied. In~\cite{6} and~\cite{7}, investigating the profile of molecules based on genetic expressions, reliable cancer classifiers are designed. The gene and/or protein changes due to certain types of cancer lead to peroxidation of cell membrane. This emits biomarkers in the blood or exhaled breath that may be detected using tailor-made cross-reactive sensors~\cite{8,9}. A high level of insulin-like growth factor or estrogen in the blood of women before menopause is one sign of breast cancer~\cite{10}. In~\cite{11}, combining engineered proteins with an appropriate detection technique is suggested to enable a new type of molecular sensor. Also in~\cite{12}, nano-bio sensors are designed and simulated for dye molecules targeting to enhance targeting efficiency.

The development of novel mathematical models and analytical approaches for disease diagnostics in the nano-scale is crucial to take advantage of nano-technology for this purpose. The mathematical modeling and simulation of cancer progression are studied in~\cite{13} and~\cite{14}, respectively. A model for nano-communications channel is proposed in~\cite{15,16}. In~\cite{17}, the noise in diffusion-based molecular communication over nano-networks is analyzed. The design of optimized molecular recognizers is studied in the biochemical noisy environment using a Bayesian cost in~\cite{18}. Such recognizers could serve as abnormality detection (AD) mechanisms by distinguishing between two molecule types, which one exists in the body on the healthy setting and the other appears only in the presence of a certain disease. In~\cite{19}, a layered architecture of molecular communication is investigated. In practical schemes for abnormality detection, one can typically identify a two-tier architecture for detection. In the first tier, the presence of abnormality is detected in the molecular nano-scale level. In the second tier, the abnormality is reported in a bigger scale to a data gathering node (DGN) in the outside world. A similar hierarchical architecture, which includes two levels of nano and micro scale messages is considered in~\cite{20} for body sensor networks. In~\cite{21}, a two-tier nano abnormality detection scheme (NADS) in which the sensor nano-machines (SNMs) have independent Poisson observations~\cite{17} is suggested and its detection performance is analyzed.

Table~\ref{Table1} presents the two tiers of abnormality detection, i.e., detection in nano-scale and micro-scale, in different methods of cancer detection based on nano-technology.
In detection of breast cancer, quantum dot bio-conjugates with targeting antibodies have been used to recognize associated molecular signatures including ERBB2 (Avian erythroblastosis oncogene B-2)~\cite{22,23}. In the second tier, this feature is recognized using long-term multiple color imaging or immune-fluorescent labeling~\cite{22,23}. In early detection of lung cancer, the increasing level of epidermal growth factor receptor (EGFR) can react at the nano-scale with injected single chain forward variable (SCFV) polypeptide with embedded Au~\cite{24}, which act as a SNM. Next, the product of this reaction may be recognized by imaging techniques for finding Au in the body~\cite{25}.



\begin{table}[h]
\caption{NADS in cancer detection using nano-technology.}
\label{Table1}
\small
\begin{tabular}{|p{4cm}|p{6cm}|p{5cm}|}
\hline
Cancer detection method                   & Detection in tier 1-NCC  & Detection in tier 2-MCC                   \\ \hline
\multirow{3}{*}   {}Nano-sized magnetic resonance imaging (MRI) contrast agents for intraoperative imaging in the context of neuro-oncological interventions~\cite{27,30}  & Gadolinium-based nano-particle~\cite{29}  & \multirow{3}{*}  {}Combined MRI with biological targeting~\cite{26} and optical detection~\cite{26,27,28}  \\ \cline{2-2}
                    & Ironoxide-based nano-particles~\cite{31,36}  &                     \\ \cline{2-2}
                    & Multiple-mode imaging contrast nano-agents &                     \\ \hline
\multirow{2}{*} {Optical detection}  & Semiconductor nano-crystals~\cite{37,38,39,40} & \multirow{2}{*} {}Optical detection \\ \cline{2-2}
                    & Quantum dots~\cite{37,38,39,40}  &                     \\ \hline
Nano scale field-effect bio-transistor                  & Silicon nano-wires~\cite{41,42} & Reporting changes in their conductance that are generated by molecular binding events on their surface                  \\ \hline
Carbon nano-technology                  & Nano-tubes have been reported as high-specificity sensors of antibody signatures of autoimmune disease~\cite{43} and of single-nucleotide polymorphisms (SNPs)~\cite{44} & Electronic biosensors                  \\ \hline
Quantum dot bio-conjugates with targeting                 & Molecular signatures including ERBB2~\cite{22,23} & Long-term multiple color
imaging, immune-fluorescent labeling
                  \\ \hline
\multirow{3}{*}{Nano-particle-based methods} & Covalently linked antibodies~\cite{45,46}  & Confocal microscopy                 \\ \cline{2-3}
                    & Fluorophore-laden silica beads have been used for the identification of leukaemia cells in blood samples~\cite{47}& Optical identification                  \\ \cline{2-3}
                    & Fluorescent nanoparticles have been used for an ultrasensitive DNA-detection system~\cite{48}
 & Fluorescence identification                  \\ \hline
\end{tabular}
\end{table}

A potential candidate for SNM in NADS is graphene-based bio-sensors, which are optimized for detecting proteins, nucleic acids, carbohydrates, or compounds generated by metabolic processes. Existing detection methods employed by these sensors include electrical, electrochemical, and photonic approaches with respect to detecting labeled (or enzyme-assisted) and label-free (or enzyme-free) probe structures~\cite{Wang}. In this context, design and analysis of a wireless nanosensor network for monitoring human lung cells using graphene based sensors are considered in~\cite{Zarepour}, where graphene antennas would be able to communicate in the terahertz band. In this case, respiration is the major process that influences the terahertz channel inside lung cells. The channel has been characterized as a two-state channel, where it periodically switches between good and bad states. It has been shown that the channel absorbs terahertz signal much faster when it is in the bad state as opposed to the good state~\cite{Zarepour}. Another reported application of in-vivo wireless network is graphene-based wireless bacteria detection on tooth enamel~\cite{Manoor}. In this case, the DGN based on terminology of~\cite{BioNanothings} is a so-called bio-cyber interface on the skin, which receives the electromagnetic signal transmitted by the graphene-based SNMs.

For the second tier of abnormality detection architecture, one may also consider the recent proposals of wireless nano sensor networks; including diffusion-based molecular communication~\cite{AkyildizNanonetworking}, medical imaging techniques \cite{31}, ultrasonic communications ~\cite{Ultrasonic},~\cite{Santagati}, optical communication using plasmonic nano-antennas~\cite{Guo} and terahertz communication techniques~\cite{Terahertz_modling}.

In this paper, a nano abnormality detection scheme (NADS) is proposed for the detection of nano-scale abnormality in a bio-molecular environment using a two-tier decision-making process. The abnormality is due to the existence of competitor cells in the said environment. The NADS includes a set of SNMs for the detection of a nano-scale abnormality over a nano-communication channel (NCC) with spatially correlated noise. The spatial correlation of noise among SNMs is motivated by the nature of bio-molecular environment in the nano-scale, and as we shall demonstrate highly influences the overall detection performance of the NADS. The SNMs communicate their decisions over a noisy micro-communication channel (MCC) to a data gathering node using micro-scale messages (MSMs). Fusing the collected signals, the DGN makes a decision and may alarm the presence of an abnormality as necessary. In Table~\ref{Table20} list of used acronyms in this paper and their definitions are discribed.

The performance analysis of the SNMs over the NCC is set up as a generalized likelihood ratio test, which quantifies the probability of false alarm and the probability of misdetection. Next, incorporating the effect of MCC, the total detection performance of NADS at the DGN is analyzed. The correlated noise in the NCC is assumed Gaussian (similar to~\cite{49,50,51,52}). In this case, the overall NADS detection performance is efficiently approximated and expressed in terms of the performances of the constituent NCCs and MCC. The presented analyses are then used to obtain the optimized concentration of SNMs in the sample for a prescribed high probability of abnormality detection and a bounded false alarm probability. Extensive numerical results are provided to quantify the effect of different design and system parameters on the NADS performance. Specifically, the effects of temporal and spatial correlation of noise at the SNMs on the detection performance are investigated.

\begin{table}[]
\centering
\caption{List of acronyms}
\label{Table20}
\begin{tabular}{|c|c|}
\hline
Acronym & Definition                         \\ \hline
AWGN    & Additive white Gaussian noise     \\ \hline
AD      & Abnormality detection              \\ \hline
DGN     & Data gathering node                \\ \hline
EGFR    & Epidermal growth factor receptor   \\ \hline
ERBB2   & Avian erythroblastosis oncogene B-2 \\ \hline
GLRT    & Generalized likelihood ratio test  \\ \hline
MAP     & Maximum a-posteriori probability   \\ \hline
MCC     & Micro communication channel        \\ \hline
MRI     & Magnetic resonance imaging         \\ \hline
MSM     & Micro-scale message                \\ \hline
NADS    & Nano abnormality detection scheme  \\ \hline
NCC     & Nano communication channel         \\ \hline
PDF     & Probability distribution function  \\ \hline
SNM     & Sensor nano-machine                \\ \hline
VTNM    & Virtual transmitter nano-machine   \\ \hline
\end{tabular}
\end{table}


The outline of this paper is as follows. In Section II, preliminaries and problem statement are presented. The communication strategy on nano and micro communication channels are described in Section III. In Section IV, the performance of NADS is evaluated analytically. Numerical results are presented in Section V. Finally, conclusions are made in Section VI.

\section{Preliminaries and Problem Statement}
In this Section, the setup of NADS and the problem statement under consideration are described. The NADS comprises of two tiers. In the first tier, each SNM detects the detection feature in nano-scale and emits a micro-scale message~\cite{Standards}. In the second tier, a DGN collects the transmitted MSMs from multiple SNMs.

The NCC models the molecular environment. In the healthy setting, no abnormality (here competitor cell) exists in the molecular environment. The molecular competitor changes the rate of binding between the molecules and the nano-receptors on the SNM or changes the number of transmitted molecules by the so-called virtual transmitter nano-machine (VTNM). This is reflected in the NCC model, with the VTNM as the transmitter and the SNMs as the receivers.

Each of the SNMs generate an MSM as it detects an abnormality. The DGN collects the MSMs over a noisy micro-communication channel. Then it decides, and declares the presence or the absence of the abnormality to the outside world. The MCC is considered an additive white Gaussian noise (AWGN) channel. Below, we continue with a detailed description of the NCC model and the detection feature.

\begin{table}
    \centering
    \caption{Effective parameters in transition probability of NCC~\cite{15}}
    \label{Table2}
    \small
    \begin{tabular}{| l | p{11cm} | l |}
    \hline
    Parameter & Parameter Description & Unit \\ \hline
    $\theta$ & Temperature & \si{\kelvin}\\ \hline
    $\chi $ & Distance between nano-transmitter and SNM & \si{\metre} \\ \hline
    $C_R$ & Concentration of nano receptors, denoted by $R$, on the SNM  & \si[per-mode=symbol]{\micro\mol\per\litre}   \\ \hline
    ${C_{A}}$ & Concentration of Molecular bit $A$, transmitted by VTNM& \si[per-mode=symbol]{\micro\mol\per\litre\per\second} \\ \hline $C_B$ &Concentration of bind-receptor, denoted by $B$, on the SNM & \si[per-mode=symbol]{\micro\mol\per\litre\per\second} \\  \hline
    ${\kappa _1}$ & Binding rate & \si[per-mode=symbol]{\micro\mol\per\litre\per\second} \\ \hline
    ${\kappa _{ - 1}}$ & Release rate & \si[per-mode=symbol]{\micro\mol\per\litre\per\second} \\ \hline
    $\kappa _0^{ - 1}$ & Zero force release & \si{\per\second} \\ \hline
    ${k_\mathit{BC}}$ & Boltzmann Constant & \si[per-mode=symbol]{\joule\per\kelvin} \\ \hline
 ${\mathit{N_{x_i}}}$  & The number of received molecule when the VTNM sends the molecular bit  $x_i \in \{A,0\} $ during time ${t_\mathit{TN}}$ at time $i$. & \si[per-mode=symbol]{\micro\mol\per\litre}  \\ \hline
 ${P_A}$ & Probability of transmission of molecular bit $A$ by the VTNM. & $ $ \\ \hline
    \end{tabular}
\end{table}

\subsection{Nano Communication Channel}
The NCC characterizes chemical reactions in the molecular environment. We consider a set of SNMs, which act as molecular receivers, injected into the biological tissue for test. We assume that the existing molecules in the molecular environment react with the receptors on SNMs. The molecules are assumed to be transmitted by a VTNM with a periodic square pulse propagation pattern. A molecular pulse $A$ (or $0$) is emitted by VTNM as $x_i$, $x_i \in\{A, 0\}$, with concentration $C_A$ (or $0$) and probability $P_A$ (or $(1-P_A)$), during time $i{t_{TN}} \le t \le i{t_{TN}} + {t_{TN}}$, $i = 0,1,...$, where $t_{TN}$ is the duration of the pulses transmitted by VTNM ~\cite{53}.
The transmitted molecules are absorbed 
at the sensor nano-machine.
If the VTNM transmits a molecular pulse $A$, the number of received molecules during the time ${t_\mathit{TN}}$ is quantified by
\begin{equation}\label{eq:1}
\mathit{N_A} = \int_0^{{t_\mathit{TN}}} {C_B\left( t \right)dt},
\end{equation}
in which $C_B\left( t \right)$ denotes the concentration of the bound receptors, in terms of ${{\mu mol} \mathord{\left/ {\vphantom {{\mu mol} {liter}}} \right. \kern-\nulldelimiterspace} {liter}}$ and is given by
\begin{equation}\label{eq:2}
C_B\left( t \right) = C_B\left( \infty  \right)\left( {1 - {e^{ - t\left( {{\kappa _{ - 1}} + {\kappa _1}{C_{A}}} \right)}}} \right), 0 \le t \le t_\mathit{TN},
\end{equation}
where $C_B\left( \infty  \right) = {{{\kappa _1}{C_{A}}C_R} \mathord{\left/
 {\vphantom {{{\kappa _1}{C_{A}}N} {\left( {{\kappa _{ - 1}} + {\kappa _1}{C_{A}}} \right)}}} \right.
 \kern-\nulldelimiterspace} {\left( {{\kappa _{ - 1}} + {\kappa _1}{C_{A}}} \right)}}$ is the steady state concentration of the bound receptors~\cite{15}. The parameters ${\kappa _1}$  and ${\kappa _{ - 1}}$, respectively are binding and release rates for the following reactions
\begin{subequations}\label{eq:3}
\begin{align}
A + R\mathop  \to \limits^{{\kappa _1}} B, \\
B\mathop  \to \limits^{{\kappa _{ - 1}}} A + R,
\end{align}
\end{subequations}
where $R$ and $B$ respectively, denote nano-receptors on the SNMs and the bound-receptors after reaction between $A$ and $R$. It is evident in \eqref{eq:2} that $C_B\left( t \right)$ is increased exponentially over time within the pulse period with concentration of $C_{A}$. After time ${t_{TN}}$, when the pulse duration ends, $C_B\left( t \right)$ is reduced as
\begin{equation}\label{eq:4}
C_B\left( t \right) = C_B\left( {{t_{TN}}} \right)\exp \left( { - {\kappa _{ - 1}}\left( {t - {t_{TN}}} \right)} \right){\rm{  }}~ {\rm{for}} ~{\rm{  }}t > {t_{TN}}.
\end{equation}
As a result, at the SNM and over the subsequent time interval, this previous pulse is reflected as follows in the receiver
\begin{equation}\label{eq:5}
{\mathit{N'_A}} = \int_0^{{t_\mathit{TN}}} {\mathit{N_A}} {e^{\left( { - {\kappa _{ - 1}}t} \right)}}dt.
\end{equation}
Obviously, we have $\mathit{N_0}=\mathit{N'_0}=0$. The rates of interaction of the molecules with the SNM receptors, ${\kappa _1}$ and ${\kappa _{-1}}$, depend on the molecular diffusion over the NCC. Hence, ${\kappa _1}$ may be influenced by such parameters as the molecular diffusion coefficient and the temperature of the environment, $\theta$~\cite{54}, and may be assessed analytically~\cite{55}. The release rate, ${\kappa _{ - 1}}$ is given by~\cite{54}
\begin{equation}\label{eq:6}
{\kappa _{ - 1}} = \kappa _{ - 1}^0{e^{{{\chi \upsilon } \mathord{\left/
 {\vphantom {{\chi \upsilon } {{k_\mathit{BC}}\theta }}} \right.
 \kern-\nulldelimiterspace} {{k_\mathit{BC}}\theta }}}},
\end{equation}
in which $\upsilon $ depends on the energy of the molecules propagated between the VTNM and the SNMs and environment factors, and $\chi$, $k_\mathit{BC}$ and $\theta$ are defined in Table~\ref{Table2}. The parameter $\kappa _{ - 1}^0$ can be obtained by matching experimental measurements, and depends on the absorption capability of molecules at the SNM~\cite{54}. Hence, it is assumed that $\kappa _{ - 1}^0$ is a variable, which depends on the properties of nano-receptors in the SNM.
\vspace{-0.2cm}

The noise of the SNM measurement is correlated over time and space. The former is due to the slow variation of SNM measurement as it models a bio-chemical reaction. The latter is due to the relatively small volume of the molecular environment in the range of $nm^3$ to $\mu m^3$. The NCC is modeled by a first order Markov model with additive noise, and as such the input of SNM $j$ at time $i$ is described by
\begin{equation}\label{eq:7}
{y_{ij}} = {{g^ + }\left( {{x_i},{\kappa _1},{\kappa _{ - 1}},{t_\mathit{TN}},{C_A},\theta } \right) + {g^ - }\left( {{x_{i - 1}},{\kappa _1},{\kappa _{ - 1}},{t_\mathit{TN}},{C_{A}},\theta } \right)}  + {\varepsilon _{ij}}.
\end{equation}
 In~\eqref{eq:7}, if the VTNM transmits the molecular bit ${x_i} \in \{A,0\}$, then
\begin{equation}\label{eq:8}
{g^ + }\left( {{x_i},{\kappa _1},{\kappa _{ - 1}},{t_\mathit{TN}},{C_{A}},\theta } \right) = \mathit{N_{x_i}},
\end{equation}
\begin{equation}\label{eq:9}
{g^ - }\left( {{x_{i - 1}},{\kappa _1},{\kappa _{ - 1}},{t_\mathit{TN}},{C_{A}},\theta } \right) = {\mathit{N'_{x_{i-1}}}}_{},
\end{equation}
indicate the number of molecules received in the current time interval from the current and previous transmissions, respectively. Also, ${\varepsilon _{ij}}$'s are jointly normal distributed with an assumed time correlation span of $p$. The temporal (normalized) covariance matrix of ${\varepsilon _{ij}}$'s is given by
\begin{equation}\label{eq:10}
{\Omega ^{\mathit{TC}}} = {\left[ {\omega _{ij}^{\mathit{TC}}} \right]_{p \times p}}.
\end{equation}
The SNM observes the nano-communications channel for a time duration of $n \ge p$.
We next consider the spatial correlation. The spatial (normalized) covariance matrix of SNM noises ${\varepsilon_{ij}}$ is given by
\begin{equation}\label{eq:12}
{\Omega ^{\mathit{SC}}} = {\left[ {\omega _{jl}^{\mathit{SC}}} \right]_{M \times M}},
\end{equation}
in which $M$ is the number of SNMs, $\omega _{jl}^{\mathit{SC}}$ is the correlation coefficient of observations of SNMs $j$ and $l$. Moreover, in this paper we assume that the space-time correlation function of SNM noises is separable~\cite{59,60}. As such, the correlation coefficient of $\varepsilon_{ij}$ and $\varepsilon_{kl}$ is given by
\begin{equation}\label{eq:13}
\frac{{E({\varepsilon _{ij}}{\varepsilon _{kl}})}}{{\sqrt {E(\varepsilon _{ij}^2)E(\varepsilon _{kl}^2)} }} = \omega _{ik}^{TC}\omega _{jl}^{SC}.
\end{equation}

The Gaussian NCC model we consider here is also justified from a molecular communication perspective. The propagation models of molecules over a diffusive molecular communication channel are widely studied in the literature~\cite{Moore,17,Pierobon,Mahfuz2,Mahfuz3,Ghavami2}. In a molecular communication system, with information encoded in the number of molecules, the number of received molecules exhibits a binomial process~\cite{Ghavami2}. When multiple emissions are considered, due to the ISI caused by the diffusion channel, previous transmissions must also be taken into account for the determination of the current symbol. This requires a summation of the binomial random variables, which is analytically hard to work with. Therefore, in the literature, two approximations of the binomial distribution are used, namely the Poisson and Gaussian approximations ~\cite{Kuran, Kilinc, Mahfuz3, Noel2}. In~\cite{Yilmaz}, it is shown that when the number of transmitted molecules increases, the Gaussian approximation provides a good model for the molecular communications channel.


\subsection{Detection Feature}
The biochemical activities of the competitor cells, e.g., cancer cells, affect the molecular environment and change its parameters~\cite{18}. We model this as an abnormality or intrusion in the molecular environment, which is to be detected as early as possible. The presence of competitor cells affects the NCC. For example, the competitor cells can react with the molecules transmitted by the VTNM. This reduces the concentration of transmitted molecules ${C_{A}}$, and hence, changes the NCC parameters or input. This variation in NCC parameters or input is used for modeling of protein identification for early cancer detection in the nano-scale~\cite{18}. Alternatively, the competitor cells may devitalize the receptors on the SNMs, change $\kappa _{ - 1}^{}$  and $\kappa _1$ on the SNM by a biochemical reaction or vary the temperature of nano-receptors on the SNMs.

In the NCC, for a given size of sample tissue and the parameters in Table~\ref{Table2}, a measurable parameter is defined as detection feature, which is to be constant during measurement. In presence of competitor cells, this parameter deviates from its normal value, that in turn is detected by the SNM. Here, we consider two scenarios although other scenarios may also be similarly considered. In the first scenario, we assume that the VTNM always sends molecular bit 0 ($P_A = 0$) in the healthy setting, and sends only molecular bit $A$ ($P_A = 1$) when an abnormality exists. In this case, the detection feature is defined as follows
\begin{equation}\label{eq:121}
\mathit{NR} ={{g^ + }\left( {{x_i},{\kappa _1},{\kappa _{ - 1}},{t_\mathit{TN}},{C_{A}},\theta } \right) + {g^ - }\left( {{x_{i - 1}},{\kappa _1},{\kappa _{ - 1}},{t_\mathit{TN}},{C_{A}},\theta } \right)} .
\end{equation}
Hence, $\mathit{NR}$ in~\eqref{eq:121} is a constant value in the healthy setting and changes to another constant value as the environmental parameters vary in the non-healthy setting.

In the second scenario, we assume that VTNM sends molecular bit $A$ with probability $P_A$ and the presence of a competitor cell in the environment can change $P_A$ and/or channel parameters. In this case, $\mathit{NR}$ in~\eqref{eq:121} is not a constant value over multiple transmissions, but its average is still so.
As such, the detection feature is defined as follows
\begin{equation}\label{eq:131}
\mathit{NR} = E\left( {{g^ + }\left( {{x_i},{\kappa _1},{\kappa _{ - 1}},{t_\mathit{TN}},{C_{A}},\theta } \right) + {g^ - }\left( {{x_{i - 1}},{\kappa _1},{\kappa _{ - 1}},{t_\mathit{TN}},{C_{A}},\theta } \right)} \right),
\end{equation}
where, the expectation (average) at the receiver is naturally computed over multiple transmission time slots, $t_{TN}$. By this definition, $\mathit{NR}$ in~\eqref{eq:131} has two distinct constant values in the healthy and non-healthy settings, and is used as an abnormality detection feature. In this case, we rewrite~\eqref{eq:7}, with a new channel output interpretation, as follows (this allows us to treat both scenarios in a common setting in the sequel)
 \begin{equation}\label{eq:171}
{y_{ij}} = E\left( {{g^ + }\left( {{x_i},{\kappa _1},{\kappa _{ - 1}},{t_\mathit{TN}},{C_{A}},\theta } \right) + {g^ - }\left( {{x_{i - 1}},{\kappa _1},{\kappa _{ - 1}},{t_\mathit{TN}},{C_{A}},\theta } \right)} \right) + {\varepsilon _{ij}}.
\end{equation}
Note that the same Gaussian model described in \eqref{eq:10}-\eqref{eq:13} for ${\varepsilon _{ij}}$ is adopted here. Obviously the model parameters may not be necessarily the same in the two mentioned scenarios. It is noteworthy that the separability of the space-time correlation function remains valid.

In both noted scenarios, the NCC is considered homogeneous and we have $E\left[ {{{\left( {{y_{ij}} - {\mathit{NR}}} \right)}^2}} \right] = E\left[ {{{\left( {{y_{il}} - \mathit{NR}} \right)}^2}} \right]$, $j,l \in \left\{ {1,...,M} \right\}$ and $i \in \left\{1,...,n\right\}$. In the healthy setting, $\mathit{NR} = \mathit{NH}$; and in presence of a competitor cell or an abnormality that affects the NCC parameters or input, $\mathit{NR}$ deviates from $\mathit{NH}$. In the sequel, we consider $y_{ij}$ as a decision variable, whose time average $\mathit{NR}$ serves as a detection feature for abnormality detection at SNM $j$.



\subsection{Problem Statement}
We consider a design optimization problem to determine the minimum required concentration of SNMs, $\bar M = {M \mathord{\left/ {\vphantom {M {vol}}} \right. \kern-\nulldelimiterspace} \mathit{vol}}$ , in the test environment for a reliable NADS, where $\mathit{vol}$ is the volume of the sample. The SNMs are typically synthesized chemical compounds that could be expensive or could create side effects if used in vivo. Hence, we wish to use them in the smallest concentration possible. A reliable NADS would identify the existence of an abnormality with sufficiently high probability, $P_D$. At the same time, when the abnormality in fact does not exist it only makes a (false) alarm with sufficiently small probability, $P_F$. The $P_D$ and $P_F$ are later analyzed in Theorem 3. The desired optimization problem in this paper is formulated as follows.

\emph{Problem}. The NADS design optimization problem is given by 
\begin{eqnarray}\label{eq:15}
&{{\bar M}^*}{\rm{ = }}\min \bar M\\ \nonumber
{\rm{   }}\rm{subject}~\rm{to}& {P_D} \ge \xi ,{P_F} \le \gamma .
\end{eqnarray}
where, $\xi $ is a constant close to unity and $\gamma $ is a constant close to zero.
As observed in Section V for given values of $\xi $ and $\gamma $, the optimized concentration of SNMs, $\bar M$, depends on type or level of abnormality, $k$.

\section{Detection Strategy over NCC and MCC}
In this Section, the detection strategy over NCC and MCC is studied. In the first Subsection, a hypothesis test is set up for the detection of competitor cells in the bio-molecular environment. Subsequently, the communication and detection strategies over the MCC are studied.

\subsection{Hypothesis Test for AD in NCC}
This test determines the functionality of the SNM over the NCC. We derive a threshold level for each SNM to alarm the presence of competitor cells by generating a micro scale message. This is accomplished such that the detection probability of each SNM over the NCC is maximized for a bounded probability of false alarm. The detection probability in terms of the false alarm probability is the basic performance characteristic of an SNM over the NCC.

The following hypothesis test is considered for the detection of a competitor cell in the molecular environment
\begin{equation}\label{eq:16}
\left\{ {\begin{array}{*{20}{c}}
{{H_0},}&{\mathit{NR} = \mathit{NH}}\\
{{H_1},}&{\mathit{NR} \ne \mathit{NH}.}
\end{array}} \right.
\end{equation}

The Gaussian assumption for the observation is motivated based on thermal noise distribution, the noise in gene expression levels~\cite{50} and the noise of biochemical systems~\cite{52}. In the sequel, the detection performance of the hypothesis test in~\eqref{eq:16} is analyzed, where we consider maximum likelihood estimate of $\mathit{NR}$ at the SNM over the observation period $n$, i.e., $\widehat {\mathit{NR}_j} = \arg \mathop {\max }\limits_{{\mathit{NR}}} {P\left( {\left. {{{\bf{y}}_j^n}} \right|{\mathit{NR}}} \right)} $, where ${{\bf{y}}_j^n} = {\left[ {{y_{1j}},{y_{2j}},...,{y_{nj}}} \right]^\dag }$, where $^\dag$ denotes the transpose operation. If we rewrite \eqref{eq:7} and \eqref{eq:171} at the receiver in terms of $\mathit{NR}$, respectively based on definition of $\mathit{NR}$ in~\eqref{eq:121} or ~\eqref{eq:131}, for SNM $j = 1,2, \ldots ,M$ and time $i = 1,2, \ldots ,n$, we have
\begin{equation}\label{eq:17}
{y_{ij}} = {\mathit{NR}} + {\varepsilon _{ij}}.
\end{equation}
Without loss of generality, we consider $n \ge p$, and define the extended temporal (normalized) covariance matrix of observations within the observation period $n$ as follows

\begin{equation}\label{eq:18}
{\Omega ^T} = {\left[ {\omega _{ij}^{\mathit{TC}}} \right]_{n \times n}} = {\left[ {\begin{array}{*{20}{c}}
{{\Omega ^{\mathit{TC}}}}& \cdots &\underline 0\\
 \vdots & \ddots & \vdots \\
\underline 0& \cdots &{{\Omega ^{\mathit{TC}}}}
\end{array}} \right]_{n \times n}}.
\end{equation}

For example with $p = 2$ and $\omega _{12}^{\mathit{TC}} = \omega _{21}^{\mathit{TC}} = \rho$ , ${\Omega ^T}$ is given by
\begin{equation}\label{eq:19}
{\Omega ^T} = {\left[ {\begin{array}{*{20}{c}}
1&\rho & \cdots &0\\
\rho &1& \cdots &0\\
 \vdots & \vdots & \ddots &\vdots \\
0&0&\rho &1
\end{array}} \right]_{n \times n}}.
\end{equation}

By this model of channel, as the status of the molecular environment departs from a healthy setting, the detection feature, $\mathit{NR}$, deviates from $\mathit{NH}$. Here, $\mathit{NR}$ deviates form $\mathit{NH}$ as follows
\begin{equation}\label{eq:14}
\mathit{NR} = \left( {1 \pm k{\sigma _{\mathit{NCC}}}} \right)\mathit{NH},
\end{equation}
in which $k$ could indicate the type or level of abnormality and $\sigma_{NCC}$ is standard deviation of noise in NCC. For $k = 0$, the molecular environment is healthy ($\mathit{NR} = \mathit{NH}$) and we assume $1 \pm k{\sigma _{\mathit{NCC}}} \ge 0$ , $k \ge 0$. A certain value of $k$ could correspond to a given progress level of a disease.

The conditional probability of observations vector ${\bf{y}}_j^n$ , given ${\mathit{NR}}$ at SNM $j$ is computed as follows, where $^\dag$ denotes the transpose operation,
\begin{equation}\label{eq:20}
\begin{array}{l}
P\left( {\left. {{{\bf{y}}_j^n}} \right|{\mathit{NR}}} \right) = P\left( {\left. {{y_{1j}},{y_{2j}}, \ldots ,{y_{nj}}} \right|{\mathit{NR}}} \right) = \\
\frac{1}{{{{\left( {2\pi } \right)}^{{n \mathord{\left/
 {\vphantom {n 2}} \right.
 \kern-\nulldelimiterspace} 2}}}\sigma _{\mathit{NCC}}^n{{\left| {{\Omega ^{\emph{T}}}} \right|}^{{1 \mathord{\left/
 {\vphantom {1 2}} \right.
 \kern-\nulldelimiterspace} 2}}}}}\exp \left( { - \frac{1}{{2\sigma _{\mathit{NCC}}^2}}{{\left( {{{\bf{y}}_j^n} - {\bf{NR}}^n} \right)}^\dag }{\Omega ^T}^{^{ - 1}}\left( {{{\bf{y}}_j^n} - {\bf{NR}}^n} \right)} \right).
\end{array}
\end{equation}
and ${\bf{NR}}^n = {\left[ {\begin{array}{*{20}{c}}{{\mathit{NR}}}&{{\mathit{NR}}}& \cdots &{{\mathit{NR}}}
\end{array}} \right]^\dag_{n \times 1}}$. Considering the Logarithm of~\eqref{eq:20}, we have
\begin{equation}\label{eq:21}
\log P\left( {\left. {{{\bf{y}}_j^n}} \right|{\mathit{NR}}} \right) =  - n\log \left( {{{\left| {{\Omega ^T}} \right|}^{{1 \mathord{\left/
 {\vphantom {1 {\left( {2n} \right)}}} \right.
 \kern-\nulldelimiterspace} {\left( {2n} \right)}}}}\sqrt {2\pi \sigma _{\mathit{NCC}}^2} } \right) - \frac{1}{{2\sigma _{\mathit{NCC}}^2}}{\left( {{{\bf{y}}_j^n} - {\bf{N}}_R^n} \right)^\dag }{\Omega ^T}^{^{ - 1}}\left( {{{\bf{y}}_j^n} - {\bf{N}}_R^n} \right).
\end{equation}
We define
\begin{equation}\label{eq:22}
{\Psi ^T} = {\left[ {\psi _{ij}^{\mathit{TC}}} \right]_{n \times n}} \buildrel \Delta \over = {\Omega ^{{T^{ - 1}}}},
\end{equation}
and rewrite~\eqref{eq:21} as follows
\begin{equation}\label{eq:23}
\log P\left( {\left. {{{\bf{y}}_j^n}} \right|{\mathit{NR}}} \right) =  - n\log \left( {{{\left| {{\Omega ^T}} \right|}^{{1 \mathord{\left/
 {\vphantom {1 {\left( {2n} \right)}}} \right.
 \kern-\nulldelimiterspace} {\left( {2n} \right)}}}}\sqrt {2\pi \sigma _{\mathit{NCC}}^2} } \right) - \frac{1}{{2\sigma _{\mathit{NCC}}^2}}\sum\limits_{l = 1}^n {\sum\limits_{i = 1}^n {\left( {{y_{lj}} - {\mathit{NR}}} \right)} } \left( {{y_{ij}} - {\mathit{NR}}} \right)\psi _{il}^{\mathit{TC}}.
\end{equation}
To maximize (23), we set its derivative with respect to $\mathit{NR}$ to zero and considering the symmetry of ${\Omega ^T}$ obtain
\begin{equation}\label{eq:24}
\widehat {\mathit{NR_j}} = {{\sum\limits_{l = 1}^n {\sum\limits_{i = 1}^n {{y_{lj}}\psi _{il}^{\mathit{TC}}} } } \mathord{\left/
 {\vphantom {{\sum\limits_{l = 1}^n {\sum\limits_{i = 1}^n {{y_{lj}}\psi _{il}^{\mathit{TC}}} } } {\sum\limits_{l = 1}^n {\sum\limits_{i = 1}^n {\psi _{il}^{\mathit{TC}}} } }}} \right.
 \kern-\nulldelimiterspace} {\sum\limits_{l = 1}^n {\sum\limits_{i = 1}^n {\psi _{il}^{\mathit{TC}}} } }}
\end{equation}
To derive the decision rule of Neyman-Pearson as in the hypothesis test of~\eqref{eq:16}, we employ the generalized likelihood ratio test (GLRT) in the next theorem due to the hypothesis test in~\eqref{eq:16} is composite test~\cite{56}.

\emph{Theorem} 1. Consider an SNM with $n$ temporally correlated Gaussian observations over the NCC. For the hypothesis test in~\eqref{eq:16}, the decision threshold and the detection probability with limited probability of false alarm, $P_F^{\mathit{NCC}} < {\eta _1}$, are given by
\begin{equation}\label{eq:25}
\left\{ {\begin{array}{*{20}{c}}
{{H_0},}&{{\mathit{NH}} - {\sigma _D}{\phi ^{ - 1}}\left( {1 - \frac{{{\eta _1}}}{2}} \right) < \widehat \mathit{NR_j} < {\mathit{NH}} + {\sigma _D}{\phi ^{ - 1}}\left( {1 - \frac{{{\eta _1}}}{2}} \right)}\\
{{H_1},}&{\begin{array}{*{20}{c}}
{\widehat \mathit{NR_j} > {\mathit{NH}} + {\sigma _D}{\phi ^{ - 1}}\left( {1 - \frac{{{\eta _1}}}{2}} \right)}\\
{\widehat \mathit{NR_j} < {\mathit{NH}} - {\sigma _D}{\phi ^{ - 1}}\left( {1 - \frac{{{\eta _1}}}{2}} \right),}
\end{array}}
\end{array}} \right.
\end{equation}
\begin{equation}\label{eq:26}
\begin{array}{l}
P_D^{\mathit{NCC}} = 1 - \\
{{\cal Q}}\left( {{{\left( { - {\sigma _D}{\phi ^{ - 1}}\left( {1 - \frac{{{\eta _1}}}{2}} \right) \mp k{\sigma _{\mathit{NCC}}}{\mathit{NH}}} \right)} \mathord{\left/
 {\vphantom {{\left( { - {\sigma _D}{\phi ^{ - 1}}\left( {1 - \frac{{{\eta _1}}}{2}} \right) \mp k{\sigma _{\mathit{NCC}}}{\mathit{NH}}} \right)} {{\sigma _D}}}} \right.
 \kern-\nulldelimiterspace} {{\sigma _D}}}} \right) + {{\cal Q}}\left( {{{\left( {{\sigma _D}{\phi ^{ - 1}}\left( {1 - \frac{{{\eta _1}}}{2}} \right) \mp k{\sigma _{\mathit{NCC}}}{\mathit{NH}}} \right)} \mathord{\left/
 {\vphantom {{\left( {{\sigma _D}{\phi ^{ - 1}}\left( {1 - \frac{{{\eta _1}}}{2}} \right) \mp k{\sigma _{\mathit{NCC}}}{\mathit{NH}}} \right)} {{\sigma _D}}}} \right.
 \kern-\nulldelimiterspace} {{\sigma _D}}}} \right),
\end{array}
\end{equation}
where ${\phi ^{ - 1}}\left( . \right)$ is the inverse function of normal cumulative distribution, $\phi$; $\cal{Q}\left( .  \right) = $1$ - \phi \left( . \right)$ is Q-functions and
\begin{equation}\label{eq:26_0}
\\\nonumber
{\sigma _D} = \sqrt {{{\left( {\sum\limits_{l = 1}^n {\sum\limits_{i = 1}^n {\psi _{il}^{\mathit{TC}}} } } \right)}^{ - 2}}\left\{ {\sum\limits_{l = 1}^n {{{\left( {\sum\limits_{i = 1}^n {\psi _{il}^{\mathit{TC}}} } \right)}^2}\sigma _{\mathit{NCC}}^2}  + 2\sum\limits_{l = 1}^{n - 1} {\sum\limits_{q = l + 1}^n {\omega _{kl}^{\mathit{TC}}\sigma _{\mathit{NCC}}^2} \left( {\sum\limits_{i = 1}^n {\psi _{iq}^{\mathit{TC}}} } \right)\left( {\sum\limits_{i = 1}^n {\psi _{il}^{\mathit{TC}}} } \right)} } \right\}}
\end{equation}
\emph{Proof}. See Appendix A. □

The probability of miss-detection for each SNM then is given by
\begin{equation}\label{eq:27}
P_M^{\mathit{NCC}} = 1 - P_D^{\mathit{NCC}}.
\end{equation}
As we shall demonstrate in Section V, a larger $n$ would enhance the performance in general. However, the level of obtained gain depends on the temporal dependency of the SNM observations. In the next Section, we study the abnormality detection and communication over the MCC.

\subsection{Detection and Communication Strategy over MCC}
The DGN receives the MSMs from the SNMs over the MCC and declares either the existence or the absence of the competitor cells in the NCC. It is assumed that the MSMs have two alphabets. If the SNM $j$ detects the competitor cells, it generates the message ${X_j} = G$, otherwise it sets ${X_j} = 0$. Replacing~\eqref{eq:24} in the decision rule of SNM $j$ in~\eqref{eq:25}, we have
\begin{equation}\label{eq:28}
{X_j} = \left\{ {\begin{array}{*{20}{c}}
{0,}&{\mathit{NH} - {\sigma _D}{\phi ^{ - 1}}\left( {1 - \frac{{{\eta _1}}}{2}} \right) < {{\sum\limits_{l = 1}^n {\sum\limits_{i = 1}^n {{y_{lj}}\psi _{il}^{\mathit{TC}}} } } \mathord{\left/
 {\vphantom {{\sum\limits_{l = 1}^n {\sum\limits_{i = 1}^n {{y_{lj}}\psi _{il}^{\mathit{TC}}} } } {\sum\limits_{l = 1}^n {\sum\limits_{i = 1}^n {\psi _{il}^{\mathit{TC}}} } }}} \right.
 \kern-\nulldelimiterspace} {\sum\limits_{l = 1}^n {\sum\limits_{i = 1}^n {\psi _{il}^{\mathit{TC}}} } }} < \mathit{NH} + {\sigma _D}{\phi ^{ - 1}}\left( {1 - \frac{{{\eta _1}}}{2}} \right)}\\
{G,}&{\begin{array}{*{20}{c}}
{{{\sum\limits_{l = 1}^n {\sum\limits_{i = 1}^n {{y_{lj}}\psi _{il}^{\mathit{TC}}} } } \mathord{\left/
 {\vphantom {{\sum\limits_{l = 1}^n {\sum\limits_{i = 1}^n {{y_{lj}}\psi _{il}^{\mathit{TC}}} } } {\sum\limits_{l = 1}^n {\sum\limits_{i = 1}^n {\psi _{il}^{\mathit{TC}}} } }}} \right.
 \kern-\nulldelimiterspace} {\sum\limits_{l = 1}^n {\sum\limits_{i = 1}^n {\psi _{il}^{\mathit{TC}}} } }} > \mathit{NH} + {\sigma _D}{\phi ^{ - 1}}\left( {1 - \frac{{{\eta _1}}}{2}} \right)}\\
{{{\sum\limits_{l = 1}^n {\sum\limits_{i = 1}^n {{y_{lj}}\psi _{il}^{\mathit{TC}}} } } \mathord{\left/
 {\vphantom {{\sum\limits_{l = 1}^n {\sum\limits_{i = 1}^n {{y_{lj}}\psi _{il}^{\mathit{TC}}} } } {\sum\limits_{l = 1}^n {\sum\limits_{i = 1}^n {\psi _{il}^{\mathit{TC}}} } }}} \right.
 \kern-\nulldelimiterspace} {\sum\limits_{l = 1}^n {\sum\limits_{i = 1}^n {\psi _{il}^{\mathit{TC}}} } }} < \mathit{NH} - {\sigma _D}{\phi ^{ - 1}}\left( {1 - \frac{{{\eta _1}}}{2}} \right),}
\end{array}}
\end{array}} \right.
\end{equation}
The probability of the events ${X_j} = G$  and ${X_j} = 0$ depends on the presence or the absence of the competitor cell. If the competitor cell is present in the molecular environment, the probability of micro-scale message is given by
\begin{equation}\label{eq:29}
p\left( {\left. {{X_j}} \right|{\mathit{NR}} \ne {\mathit{NH}}} \right) = \left\{ {\begin{array}{*{20}{c}}
{1 - P_D^{\mathit{NCC}}}&{{X_j} = 0}\\
{P_D^{\mathit{NCC}}}&{{X_j} = G.}
\end{array}} \right.
\end{equation}
If the competitor cell is not present in the environment the probability of micro-scale message is given by
\begin{equation}\label{eq:30}
p\left( {\left. {{X_j}} \right|\mathit{NR} = \mathit{NH}} \right) = \left\{ {\begin{array}{*{20}{c}}
{1 - P_F^{\mathit{NCC}}}&{{X_j} = 0}\\
{P_F^{\mathit{NCC}}}&{{X_j} = G,}
\end{array}} \right.
\end{equation}
The signal received at the DGN through the AWGN MCC then is given by
\begin{equation}\label{eq:31}
V = \sum\limits_{j = 1}^M {{X_j}}  + {\varepsilon _{DGN}},
\end{equation}
where, ${\varepsilon _{DGN}} \sim {{\cal N}}\left( {0,\sigma _{\mathit{MCC}}^2} \right)$, and $\sigma _{\mathit{MCC}}^2$  is the MCC noise variance. We set up the following hypothesis test at the DGN,
\begin{equation}\label{eq:32}
\left\{ {\begin{array}{*{20}{c}}
{{H_0}}&{\sum\limits_{j = 1}^M {{X_j}}  < G}\\
{{H_1}}&{\sum\limits_{j = 1}^M {{X_j}}  \ge G.}
\end{array}} \right.
\end{equation}
This fusion rule is known as the OR-rule~\cite{57}. The hypothesis ${H_1}$(${H_0}$) is declared if at least one (none) of the SNMs transmits the MSM $G$, stating that the abnormality exists (does not exist) in the bio-molecular environment. In the next Section, the NADS performance is analyzed when the SNMs observations are spatially, and temporally correlated.

\section{NADS Performance Analysis}
In this Section, the performance of NADS is analyzed and two closed-form expressions for the probabilities of detection and false alarm are derived. Then, in the next Subsection, a computationally efficient formulae is derived for performance of NADS.

\subsection{Exact Performance Analysis}
\begin{figure}
        \centering
        \begin{subfigure}[b]{0.5\textwidth}
                \includegraphics[width=\textwidth]{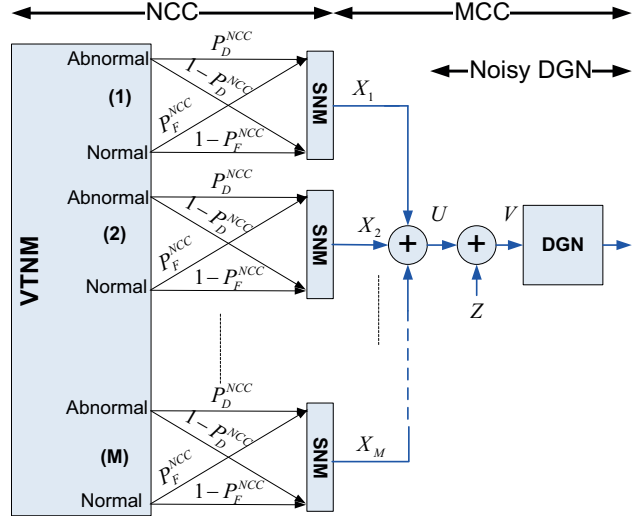}
        \end{subfigure}
        ~ 
        \caption{Modeling of communication channels between VTNM, SNMs and the DGN.}
        \label{fig:2}
\end{figure}

In this Subsection, the NADS performance is quantified using \emph{Theorem} 1 on the NCC performance and considering the communication of SNMs over the MCC as discussed in Section III. Fig.~\ref{fig:2} shows the modeling of the communication channels between the VTNM, the SNMs and the DGN. The NADS is composed of a broadcast channel with a common message followed by a Gaussian multiple access channel and an OR fusion rule.

Considering Fig.~\ref{fig:2} and the communication of SNMs over the MCC, we have
\begin{equation}\label{eq:33}
U = \sum\nolimits_{j = 1}^M {{X_j}} .
\end{equation}
The event of abnormality detection alarm at the SNM $j$, when the abnormality truly exists, is denoted by ${D_j}$  and its complementary event is denoted by ${D'_j}$. In this case, we consider ${Q_D} = P\left( {\left. {U \ge G} \right|{\mathit{NR}} \ne {\mathit{NH}}} \right)$. Considering the spatially correlated observations of SNMs and using the OR-fusion rule, ${Q_D}$ can be writen as follows
\begin{equation}\label{eq:34}
{Q_D} = 1 - \Pr \left\{ {\bigcap\nolimits_{j = 1}^M {{{D'}_j}} } \right\}.
\end{equation}
 The SNM $j$ alarms an abnormality over the NCC depending on its decision variable $\widehat {\mathit{NR_j}}$ in~\eqref{eq:24}. Hence, to quantify the probability in~\eqref{eq:34}, we need to derive the PDF of ${\widehat {{\bf{NR}}}^M} = {[\begin{array}{*{20}{c}}
{\widehat {\mathit{NR}}}_1&{\widehat {\mathit{NR}}}_2& \cdots &{\widehat {\mathit{NR}}}_M
\end{array}]^\dag }$. The next lemma serves this purpose.

\emph{Lemma }1. The decision variables of SNMs ${\widehat {{\bf{NR}}}^M} = $ ${[\begin{array}{*{20}{c}}
{\widehat {\mathit{NR}}}_1&{\widehat {\mathit{NR}}}_2& \cdots &{\widehat {\mathit{NR}}}_M
\end{array}]^\dag }$   are jointly Gaussian with mean ${\mathit{NR}}$ and (normalized) covariance matrix ${\Omega ^{\mathit{SC}}}$  in~\eqref{eq:12}.

\emph{Proof}. See Appendix B.

Using \emph{Lemma} 1 and noting the decision region in~\eqref{eq:25}, we have
\begin{equation}\label{eq:35}
\begin{array}{l}
{Q_D} = 1 - \int\limits_{{\mathit{NH}} - \tau '}^{{\mathit{NH}} + \tau '} {...\int\limits_{{\mathit{NH}} - \tau '}^{{\mathit{NH}} + \tau '} {} } \\
\frac{1}{{{{\left( {2\pi } \right)}^{{M \mathord{\left/
 {\vphantom {M 2}} \right.
 \kern-\nulldelimiterspace} 2}}}\sigma _D^M{{\left| {{\Omega ^{\mathit{SC}}}} \right|}^{{1 \mathord{\left/
 {\vphantom {1 2}} \right.
 \kern-\nulldelimiterspace} 2}}}}}\exp \left( { - \frac{1}{{2\sigma _D^2}}{{\left( {{\widehat {{\bf{NR}}}^M} - {\bf{NR}}^{M}} \right)}^\dag }{\Omega ^{\mathit{SC}}}^{^{ - 1}}\left( {{\widehat {{\bf{NR}}}^M} - {\bf{NR}}^{M}} \right)} \right)d\widehat \mathit{{NR}}_1 \ldots d\widehat \mathit{{NR}}_M
\end{array}
\end{equation}
where ${\bf{NR}}^{M} = {\left[ {\begin{array}{*{20}{c}}
{{\mathit{NR}}}&{{\mathit{NR}}}& \cdots &{{\mathit{NR}}}
\end{array}} \right]_{1 \times M}}$ . In a similar manner, we consider $E_j$  as the false alarm event that SNM $j$ alarms an abnormality, when it does not exist in reality. The complementary event is denoted by ${E'_j}$. In this case, we consider ${Q_F} = P\left( {\left. {U \ge G} \right|{\mathit{NR}} = {\mathit{NH}}} \right)$. Considering the spatially correlated observations of SNMs and using the OR-fusion rule, ${Q_F}$ can be rewritten as follows
\begin{equation}\label{eq:36}
{Q_F} = 1 - \Pr \left\{ {\bigcap\nolimits_{j = 1}^M {{{E'}_j}} } \right\}
\end{equation}
 Using \emph{Lemma} 1 and noting the decision region in~\eqref{eq:25}, we have
 \begin{equation}\label{eq:37}
 \begin{array}{l}
{Q_F} = 1 - \int\limits_{{\mathit{NH}} - \tau '}^{{\mathit{NH}} + \tau '} {...\int\limits_{{\mathit{NH}} - \tau '}^{{\mathit{NH}} + \tau '} {} } \\
\frac{1}{{{{\left( {2\pi } \right)}^{{M \mathord{\left/
 {\vphantom {M 2}} \right.
 \kern-\nulldelimiterspace} 2}}}\sigma _D^M{{\left| {{\Omega ^{\mathit{SC}}}} \right|}^{{1 \mathord{\left/
 {\vphantom {1 2}} \right.
 \kern-\nulldelimiterspace} 2}}}}}\exp \left( { - \frac{1}{{2\sigma _D^2}}{{\left( {{\widehat {{\bf{NR}}}^M} - {\bf{NH}}^{M}} \right)}^\dag }{\Omega ^{\mathit{SC}}}^{^{ - 1}}\left( {{\widehat {{\bf{NR}}}^M} - {\bf{NH}}^{M}} \right)} \right)d\widehat \mathit{NR}_1 \ldots d\widehat \mathit{NR}_M
\end{array}
\end{equation}
where ${\bf{NH}^M} = {\left[ {\begin{array}{*{20}{c}}
{\mathit{NH}}&{\mathit{NH}}& \cdots &{\mathit{NH}}
\end{array}} \right]_{1 \times M}}$ . If the observations of different SNMs are spatially independent, ${\Omega ^{\mathit{SC}}}$  is diagonal and~\eqref{eq:35} and~\eqref{eq:37} are simplified as follows~\cite{57},
\vspace{-0.5\baselineskip}
\begin{equation}\label{eq:38}
{Q_D} = 1 - {(1 - P_D^{\mathit{NCC}})^M}
\end{equation}
\vspace{-1\baselineskip}
\begin{equation}\label{eq:39}
{Q_F} = 1 - {(1 - P_F^{\mathit{NCC}})^M}
\end{equation}
where $P_F^{\mathit{NCC}} = {\eta _1}$  and $P_M^{\mathit{NCC}}$ is defined in~\eqref{eq:27}.

At the DGN with the OR-rule, we are facing a channel with binary outputs. However, the input to the DGN is a noisy version of $U$, i.e., $V$, which is the basis for the decision on the possible presence of abnormality. The next theorem presents the corresponding decision region at the DGN based on maximum a-posteriori probability (MAP) rule. This is motivated to obtain a point estimate of the unobserved quantity of presence or non-presence of abnormality based on DGN observations.

\emph{Theorem} 2. The decision region at the DGN based on MAP rule is given by
\vspace{-0.5\baselineskip}
\begin{equation}\label{eq:40}
\left\{ \begin{array}{l}
{H_0}:V < {V^{\mathit{THR}}}\\
{H_1}:V > {V^{\mathit{THR}}}
\end{array} \right.
\end{equation}
where, $V^{\mathit{THR}}$ is the minimum value of  $V$, satisfying the following inequality,
\vspace{-0.5\baselineskip}
\begin{equation}\label{eq:41}
\begin{array}{l}
\left( {\left( {1 - {Q_F}} \right)P({H_0}) - \left( {1 - {Q_D}} \right)P({H_1})} \right)\exp \left( {\frac{-{{V^2}}}{{2\sigma _{\mathit{MCC}}^2}}} \right) + \\
\left( {{Q_F}P({H_0}) - {Q_D}P({H_1})} \right)\frac{1}{{\left( {1 - {p_0}} \right)}}\sum\limits_{l = 1}^M {{p_l}\exp \left( {\frac{-{{{\left( {V - lG} \right)}^2}}}{{2\sigma _{\mathit{MCC}}^2}}} \right)} \mathop {\mathop  > \limits^{{H_0}} }\limits_ <  0.
\end{array}
\end{equation}
\vspace{-0.5\baselineskip}
In~\eqref{eq:41}, ${p_l} = \Pr \left\{ {U = lG} \right\}$   for $l \in \left\{ {0, \ldots ,M} \right\}$, is given by
\begin{equation}\label{eq:42}
{p_l} = P\left( {{H_1}} \right){{p'}_l} + P\left( {{H_0}} \right){{p''}_l}
\end{equation}
where $p'_l$ and $p''_l$ are given by
\begin{equation}\label{eq:45}
\begin{array}{l}
{{p'}_l} = \Pr \left\{ {\left. {U = lG} \right|{H_1}} \right\}=\left( {\begin{array}{*{20}{c}}
M\\
l
\end{array}} \right)\underbrace {\int_A {...} \int_A {} }_l\underbrace {\int_{{A^C}} {...} \int_{{A^C}} {} }_{M - l}\\
\frac{1}{{{{\left( {2\pi } \right)}^{{M \mathord{\left/
 {\vphantom {M 2}} \right.
 \kern-\nulldelimiterspace} 2}}}\sigma _D^M{{\left| {{\Omega ^{\mathit{SC}}}} \right|}^{{1 \mathord{\left/
 {\vphantom {1 2}} \right.
 \kern-\nulldelimiterspace} 2}}}}}\exp \left( { - \frac{1}{{2\sigma _D^2}}{{\left( {{\widehat {{\bf{NR}}}^M} - {\bf{NR}}^{M}} \right)}^\dag }{\Omega ^{\mathit{SC}}}^{^{ - 1}}\left( {{\widehat {{\bf{NR}}}^M} - {\bf{NR}}^{M}} \right)} \right)d\widehat {\mathit{NR}}_1 \ldots d\widehat {\mathit{NR}}_M
\end{array}
\end{equation}
\vspace{-0.7\baselineskip}
\begin{equation}\label{eq:46}
\begin{array}{l}
{{p''}_l} = \Pr \left\{ {\left. {U = lG} \right|{H_0}} \right\}= \left( {\begin{array}{*{20}{c}}
M\\
l
\end{array}} \right)\underbrace {\int_A {...} \int_A {} }_l\underbrace {\int_{{A^C}} {...} \int_{{A^C}} {} }_{M - l}\\
\frac{1}{{{{\left( {2\pi } \right)}^{{M \mathord{\left/
 {\vphantom {M 2}} \right.
 \kern-\nulldelimiterspace} 2}}}\sigma _D^M{{\left| {{\Omega ^{\mathit{SC}}}} \right|}^{{1 \mathord{\left/
 {\vphantom {1 2}} \right.
 \kern-\nulldelimiterspace} 2}}}}}\exp \left( { - \frac{1}{{2\sigma _D^2}}{{\left( {{\widehat {{\bf{NR}}}^M} - {\bf{NH}}^{M}} \right)}^\dag }{\Omega ^{\mathit{SC}}}^{^{ - 1}}\left( {{\widehat {{\bf{NR}}}^M} - {\bf{NH}}^{M}} \right)} \right)d\widehat {\mathit{NR}}_1 \ldots d\widehat \mathit{NR}_M.
\end{array}
\end{equation}
$\int_A \bullet  = \int_{ - \infty }^{{\mathit{NH}} - \tau '} \bullet  + \int_{{\mathit{NH}} + \tau '}^\infty \bullet$ and $\int_{{A^C}}  \bullet   = \int_{NH - \tau '}^{NH + \tau '}  \bullet$,

\emph{Proof}. See Appendix C.

Hence, the next theorem quantifies the NADS performance.

\emph{Theorem} 3. The probabilities of detection and false alarm of NADS are given by
\begin{equation}\label{eq:43}
{P_D} = {{\cal Q}}\left( {\frac{{{V^{\mathit{THR}}}}}{{{\sigma _{\mathit{MCC}}}}}} \right)\left( {1 - {Q_D}} \right) + \frac{{\sum\nolimits_{l = 1}^M {{{\cal Q}}\left( {\frac{{{V^{\mathit{THR}}} - lG}}{{{\sigma _{\mathit{MCC}}}}}} \right){{p'}_l}} }}{{1 - {{p'}_0}}}{Q_D}
\end{equation}
\vspace{-1\baselineskip}
\begin{equation}\label{eq:44}
{P_F} = {{\cal Q}}\left( {\frac{{{V^{\mathit{THR}}}}}{{{\sigma _{\mathit{MCC}}}}}} \right)\left( {1 - {Q_F}} \right) + \frac{{\sum\nolimits_{l = 1}^M {{{\cal Q}}\left( {\frac{{{V^{\mathit{THR}}} - lG}}{{{\sigma _{\mathit{MCC}}}}}} \right){{p''}_l}} }}{{1 - {{p''}_0}}}{Q_F}
\end{equation}
\emph{Proof}. The proof is provided in Appendix D.


If the noise of SNMs are considered spatially independent, the next corollary presents the NADS probability of detection and false alarm.

\emph{Corollary} 1. The probability of detection and false alarm of NADS for spatially independent NCCs are given by
\begin{equation}\label{eq:47}
{P_D} = {{\cal Q}}\left( {\frac{{{V^{\mathit{THR}}}}}{{{\sigma _{\mathit{MCC}}}}}} \right){(1 - P_D^{\mathit{NCC}})^M} + \frac{{\sum\nolimits_{l = 1}^M {Q\left( {\frac{{{V^{\mathit{THR}}} - lG}}{{{\sigma _{\mathit{MCC}}}}}} \right){{p'}_l}} }}{{1 - {{p'}_0}}}\left( {1 - {{(1 - P_D^{\mathit{NCC}})}^M}} \right)
\end{equation}
\begin{equation}\label{eq:48}
{P_F} = {{\cal Q}}\left( {\frac{{{V^{\mathit{THR}}}}}{{{\sigma _{\mathit{MCC}}}}}} \right){(1 - P_F^{\mathit{NCC}})^M} + \frac{{\sum\nolimits_{l = 1}^M {Q\left( {\frac{{{V^{\mathit{THR}}} - lG}}{{{\sigma _{\mathit{MCC}}}}}} \right){{p''}_l}} }}{{1 - {{p''}_0}}}\left( {1 - {{(1 - P_F^{\mathit{NCC}})}^M}} \right)
\end{equation}
where
\begin{equation}\label{eq:49}
{p'_l} = \left( {\begin{array}{*{20}{c}}
M\\
l
\end{array}} \right){\left( {1 - P_D^{\mathit{NCC}}} \right)^{M - l}}{\left( {P_D^{\mathit{NCC}}} \right)^l}
\end{equation}
\begin{equation}\label{eq:50}
{p''_l} = \left( {\begin{array}{*{20}{c}}
M\\
l
\end{array}} \right){\left( {1 - P_F^{\mathit{NCC}}} \right)^{M - l}}{\left( {P_F^{\mathit{NCC}}} \right)^l}.
\end{equation}

 \emph{Remark.} 1. The analysis in Theorem 3 relies on an OR rule (1 out of $M$ rule). This can be extended to the case with  $m$  out of $M$ rule at the DGN. Specifically, following similar steps, it is straight forward to show that the probabilities of detection and false alarm  are given by
\begin{equation}\label{eq:440}
P_D = \sum_{l=0}^{M} p'_l {\cal{Q}} \left( \frac {V^{\mathit{THR}}-lG}{\sigma_{MCC}}\right)
\end{equation}
\begin{equation}\label{eq:4400}
P_F = \sum_{l=0}^{M} p''_l {\cal{Q}} \left( \frac {V^{\mathit{THR}}-lG}{\sigma_{MCC}}\right).
\end{equation}
where $V^{\mathit{THR}}$ is smallest value of $V$ which satisfy the next inequality
\begin{align}
\nonumber
&\frac{{\left( {\left( {1 - {\check Q_F}} \right)P\left( {{H_0}} \right) - \left( {1 - {\check Q_D}} \right)P\left( {{H_1}} \right)} \right)}}{{\sqrt {2\pi \sigma _{MCC}^2} \sum\limits_{l = 0}^{m - 1} {{p_l}} }}\sum\limits_{l = 0}^{m - 1} {{p_l}} \exp \left( { - \frac{{{{\left( {V - lG} \right)}^2}}}{{2\sigma _{MCC}^2}}} \right) + \\
&\frac{{\left( {{\check Q_F}P\left( {{H_0}} \right) - {\check Q_D}P\left( {{H_1}} \right)} \right)}}{{\sqrt {2\pi \sigma _{MCC}^2} \sum\limits_{l = k}^M {{p_l}} }}\sum\limits_{l = m}^M {{p_l}} \exp \left( { - \frac{{{{\left( {V - lG} \right)}^2}}}{{2\sigma _{MCC}^2}}} \right)\mathop  < \limits^{\mathop  > \limits^{{H_0}} } 0
\end{align}
where $\check Q_D = \sum_{l=m}^{M} p'_l$ and $\check Q_F = \sum_{l=m}^{M} p''_l$. It is obvious that $V^{\rm THR}$   explicitly depends on $m$.  Indeed, for the case of $m = 1$ replacing   $Q_D$ and $Q_F$  in terms of $p'_l$  and $p''_l$  in~\eqref{eq:45} and~\eqref{eq:46} leads to~\eqref{eq:440} and~\eqref{eq:4400}. Our experiments (not reported here) reveal that both $P_D$  and $P_F$  reduce as $m$  increases beyond one. Assuming SNMs have small  $P_F^{\mathit{NCC}}$ and $P_D^{\mathit{NCC}}$ , and considering our application of early disease detection, in the sequel  we focus on the 1 out $M$  rule and aim at improving the probability of detection, with a small and acceptable probability of false alarm. As elaborated, similar analysis can be carried out for the case of $m$ out of $M$ rule.



\subsection{Computationally Efficient Performance Assessment}
Performance evaluation of NADS based on the analyses in Theorem 2 and equations ~\eqref{eq:42},~\eqref{eq:45} and~\eqref{eq:46} is computationally challenging in general, due to the multiple nested integrals involved (especially for large number of SNMs). Therefore, in the next lemma we present approximations that enable more computationally efficient solutions.

\emph{Lemma }2. The probabilities ${p'_l}$, ${p''_l}$  and ${p_l}$  in~\eqref{eq:42},~\eqref{eq:45} and~\eqref{eq:46} are approximated by
\begin{equation}\label{eq:51}
{{p'}_l} \approx \left( {\begin{array}{*{20}{c}}
M\\
l
\end{array}} \right){\left( {1 - P_D^{NCC}} \right)^{\alpha \frac{{M - l}}{M}\left( {{{\left[ {\bf{1}} \right]}^\dag }{\Omega ^{S{C^{ - 1}}}}\left[ {\bf{1}} \right]} \right)}}{\left( {P_D^{NCC}} \right)^{\frac{{\alpha l}}{M}\left( {{{\left[ {\bf{1}} \right]}^\dag }{\Omega ^{S{C^{ - 1}}}}\left[ {\bf{1}} \right]} \right)}},
\end{equation}
\begin{equation}\label{eq:52}
{{p''}_l} \approx \left( {\begin{array}{*{20}{c}}
M\\
l
\end{array}} \right){\left( {1 - P_F^{NCC}} \right)^{\alpha \frac{{M - l}}{M}\left( {{{\left[ {\bf{1}} \right]}^\dag }{\Omega ^{S{C^{ - 1}}}}\left[ {\bf{1}} \right]} \right)}}{\left( {P_F^{NCC}} \right)^{\frac{{\alpha l}}{M}\left( {{{\left[ {\bf{1}} \right]}^\dag }{\Omega ^{S{C^{ - 1}}}}\left[ {\bf{1}} \right]} \right)}},
\end{equation}
\begin{equation}\label{eq:53}
{\tilde p_l} = P\left( {{H_1}} \right){\tilde p'_l} + P\left( {{H_0}} \right)\tilde p''.
\end{equation}
where $\left[ {\bf{1}} \right]  = {\left[ {1,...,1} \right]^\dag_{M \times 1}}$ and $\alpha$ is a fitting parameter.

\emph{Proof}. The proof is provided in Appendix E.

Since the DGN uses an OR-fusion rule, we have ${Q_D} = 1 - {p'_0}$  and ${Q_F} = 1 - {p''_0}$. Using~\eqref{eq:51} and~\eqref{eq:52} with $l = 0$, we can approximate ${Q_D}$ and $Q_F$ in~\eqref{eq:35} and~\eqref{eq:37} for their efficient computation as follows:
\begin{equation}\label{eq:54}
{\tilde Q_D} = 1 - {(1 - P_D^{\mathit{NCC}})^{\alpha {{\left[ {\bf{1}} \right]}^\dag }{\Omega ^{\mathit{SC}}}^{^{ - 1}}\left[ {\bf{1}} \right]}}
\end{equation}
\begin{equation}\label{eq:55}
{\tilde Q_F} = 1 - {(1 - P_F^{\mathit{NCC}})^{\alpha {{\left[ {\bf{1}} \right]}^\dag }{\Omega ^{\mathit{SC}}}^{^{ - 1}}\left[ {\bf{1}} \right]}}
\end{equation}
Using the results of Lemma 2 and Corollary 1, computationally efficient expressions for $P_D$ and $P_F$ are obtained by replacing $p_l$, $p'_l$, $p''_l$, $Q_D$ and $Q_F$ with ${\tilde p_l}$, ${\tilde p'_l}$, ${\tilde p''_l}$, ${\tilde Q_D}$  and ${\tilde Q_F}$ , respectively.

\section{Numerical Results}
In this Section, we present numerical results and assess the performance of the proposed NADS. In addition, the effects of different parameters including the temporal and spatial correlations of the SNM noise are studied. In the experiments of this Section, we assume that the observations of SNMs are temporally correlated by the correlation matrix in~\eqref{eq:19} and the spatial correlation matrix is ${\Omega ^\mathit{\mathit{SC}}} = {[\omega _{_{ij}}^\mathit{\mathit{SC}}]_{M \times M}},$ and $\omega _{_{ij}}^\mathit{\mathit{SC}} = \left( {{1 \mathord{\left/
 {\vphantom {1 4}} \right. \kern-\nulldelimiterspace} 4}} \right){}^{\left| {i - j} \right|}$. We also consider the volume of the sample size at 1000 $nm^3$. Table~\ref{Table3} presents the parameters of the numerical experiments.

\begin{figure}[t!]
        \vspace{-1.5\baselineskip}
        \centering
                \includegraphics[width=0.49\textwidth]{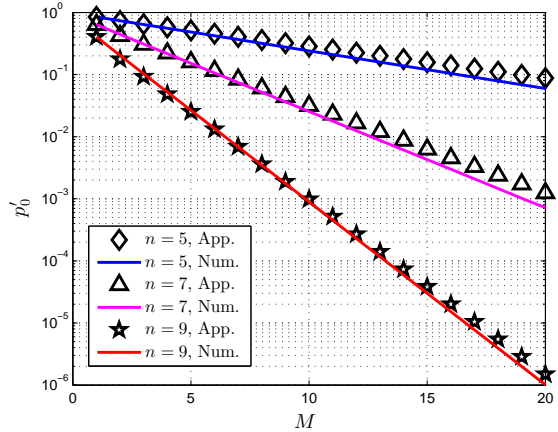}

        \vspace{-0.5\baselineskip}
        \caption {The probability of receiving no microscale messages at DGN over the spatially and temporally correlated noise of SNMs in presence of abnormality (Solid lines: ${p'_0}$ in~\eqref{eq:45}) and approximation of ${p'_0}$ (Markers: ${\tilde p'_0}$ in~\eqref{eq:51}), in terms of $M$ for different values of $n$ with $\rho  = 0.1$.} \label{fig:3}
\end{figure}

Fig.~\ref{fig:3} shows the probability of receiving no microscale messages at the DGN for spatially correlated noise of SNMs in presence of abnormality (  in ~\eqref{eq:45}) and its approximation (${p'_0}$ in~\eqref{eq:51}) in terms of the number of SNMs, $M$, for different values of observation time, $n$. As evident the approximate expression ${\tilde p'_0}$ matches the analysis ${p'_0}$ reasonably well for the selected $\alpha  = 1.2$. As such in the subsequent numerical results, we set $\alpha  = 1.2$ when using the approximations.

\begin{figure}[t!]
        \vspace{-1.5\baselineskip}
        \centering
        \begin{subfigure}[b]{0.49\textwidth}
                \caption{}
                \includegraphics[width=\textwidth]{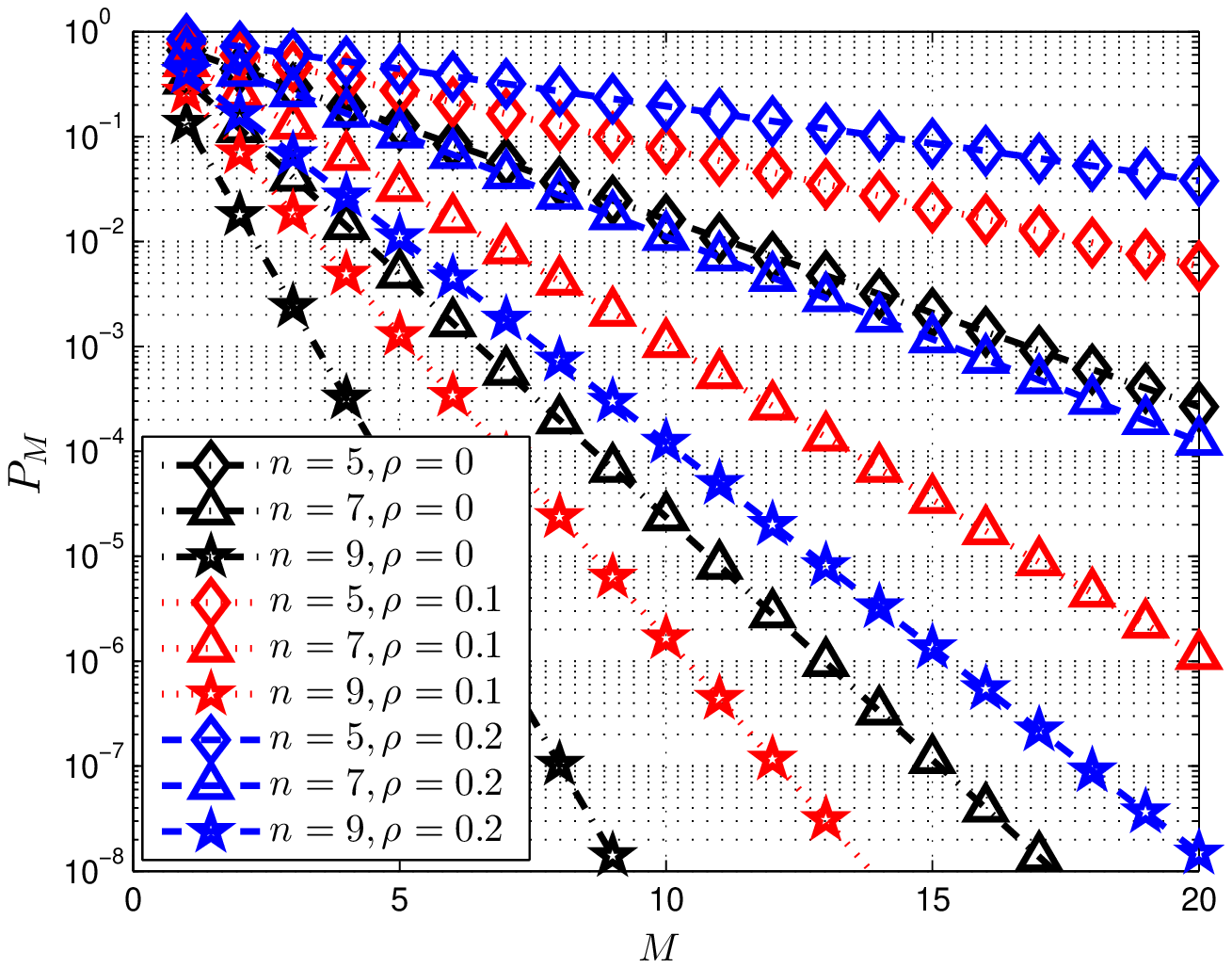}
                \label{fig:4}
        \end{subfigure}%
        \begin{subfigure}[b]{0.49\textwidth}
                \caption{}
                \includegraphics[width=\textwidth]{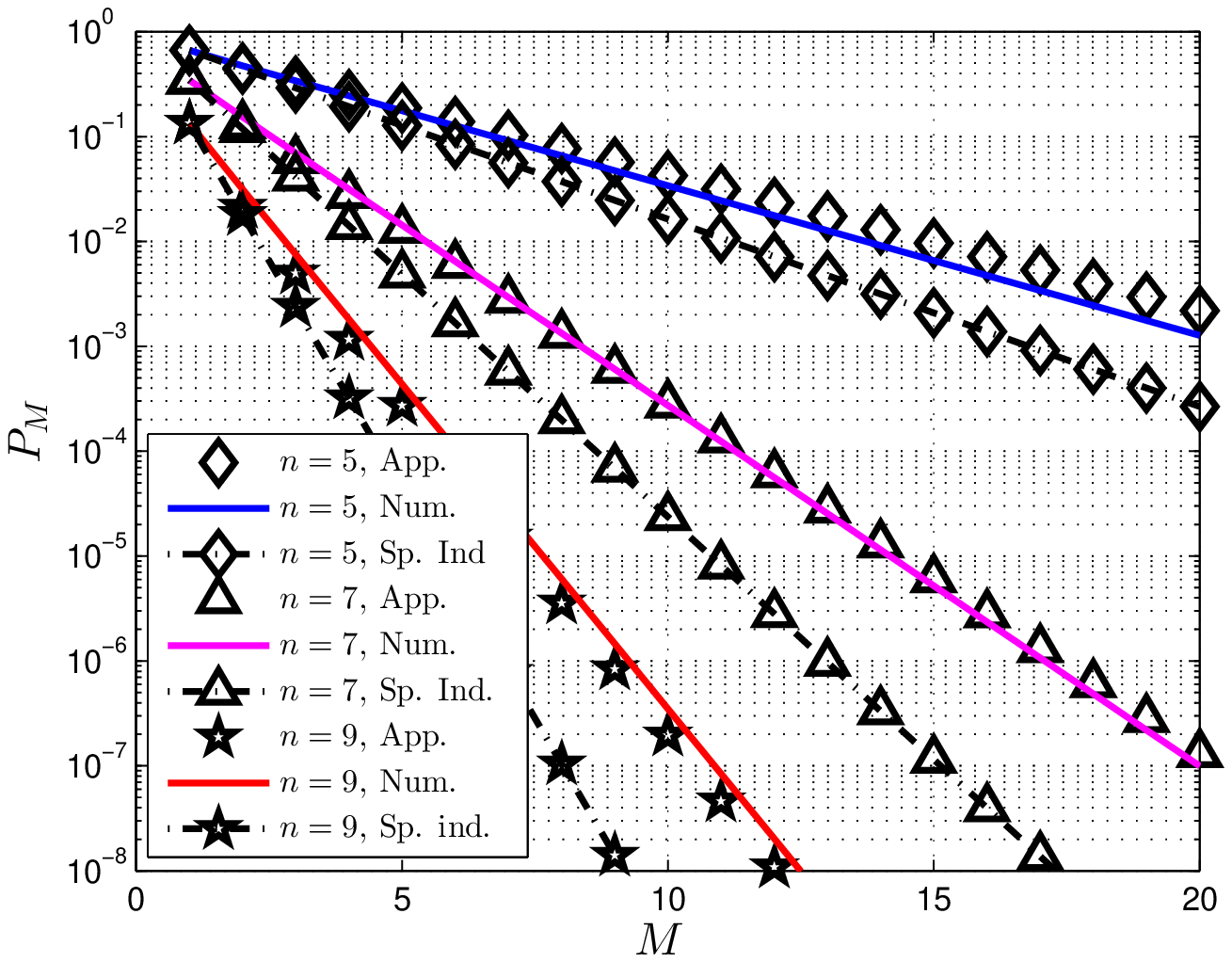}
                \label{fig:5}
        \end{subfigure} \vspace{-2em}\\
        \begin{subfigure}[b]{0.49\textwidth}
                \caption{}
                \includegraphics[width=\textwidth]{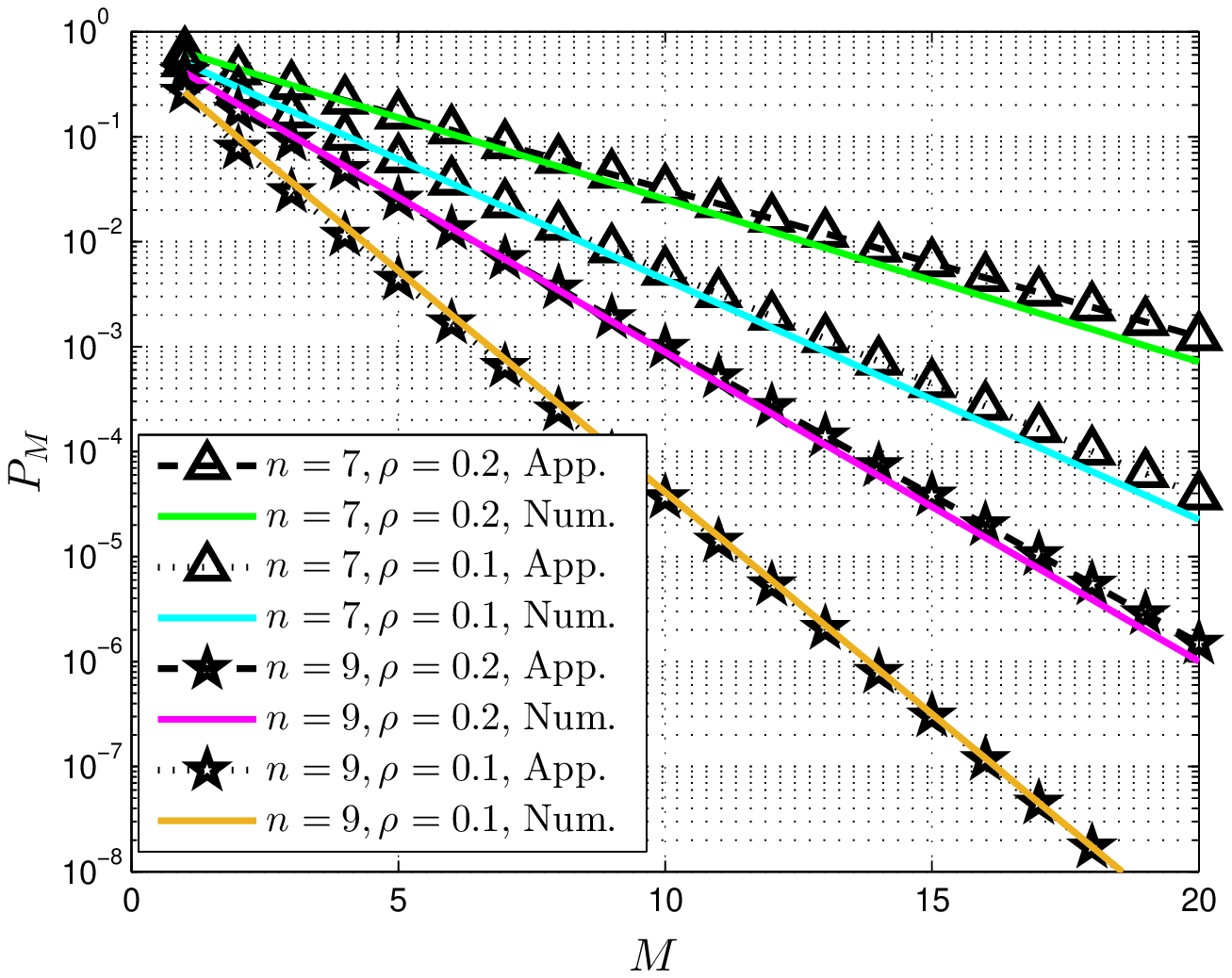}
                \label{fig:6}
        \end{subfigure}
        \begin{subfigure}[b]{0.49\textwidth}
                \caption{}
                \includegraphics[width=\textwidth]{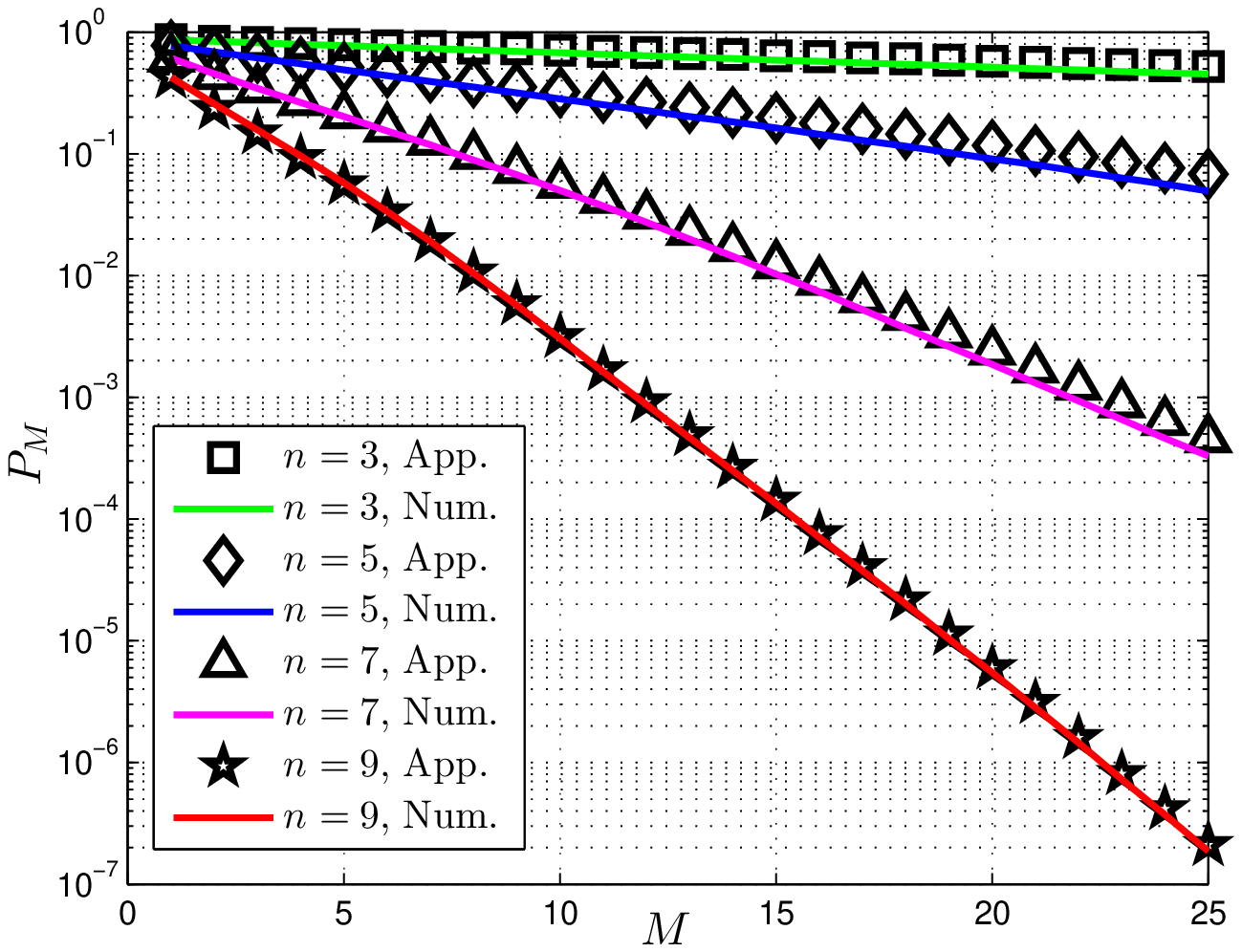}
                \label{fig:7}
        \end{subfigure}%
        \vspace{-1.5\baselineskip}
        \caption {(a) $P_M$ vs. $M$ for different values of $n$ and $\rho$  in spatially independent and temporally correlated scenario (dashed-dotted $\rho  = 0$, dotted lines $\rho  = 0.1$ , dashed lines $\rho  = 0.2$ ) and ${\sigma _{\mathit{MCC}}} = 0.1$. (b) $P_M$ vs. $M$ for different values of $n$ in spatially independent/correlated and temporally independent scenario, $\rho = 0$  and $\sigma_{\mathit{MCC}} = 0.1$. (c) $P_M$ vs. $M$ for different values of $n$ and $\rho$ in spatially and temporally correlated scenario. (Dotted lines $\rho  = 0.1$, dashed lines $\rho  = 0.2$)  and ${\sigma _{\mathit{MCC}}} = 0.1$. (d) $P_M$ in terms of $M$ for different values of $n$ in spatially and temporally correlated scenario, for $\rho = 0.2$  and $\sigma_{\mathit{MCC}} = 0.4$ . } \label{fig:P_M}
\end{figure}

\begin{figure}[t!]
        \vspace{-1.5\baselineskip}
        \centering
        \begin{subfigure}[b]{0.49\textwidth}
                \caption{}
                \includegraphics[width=\textwidth]{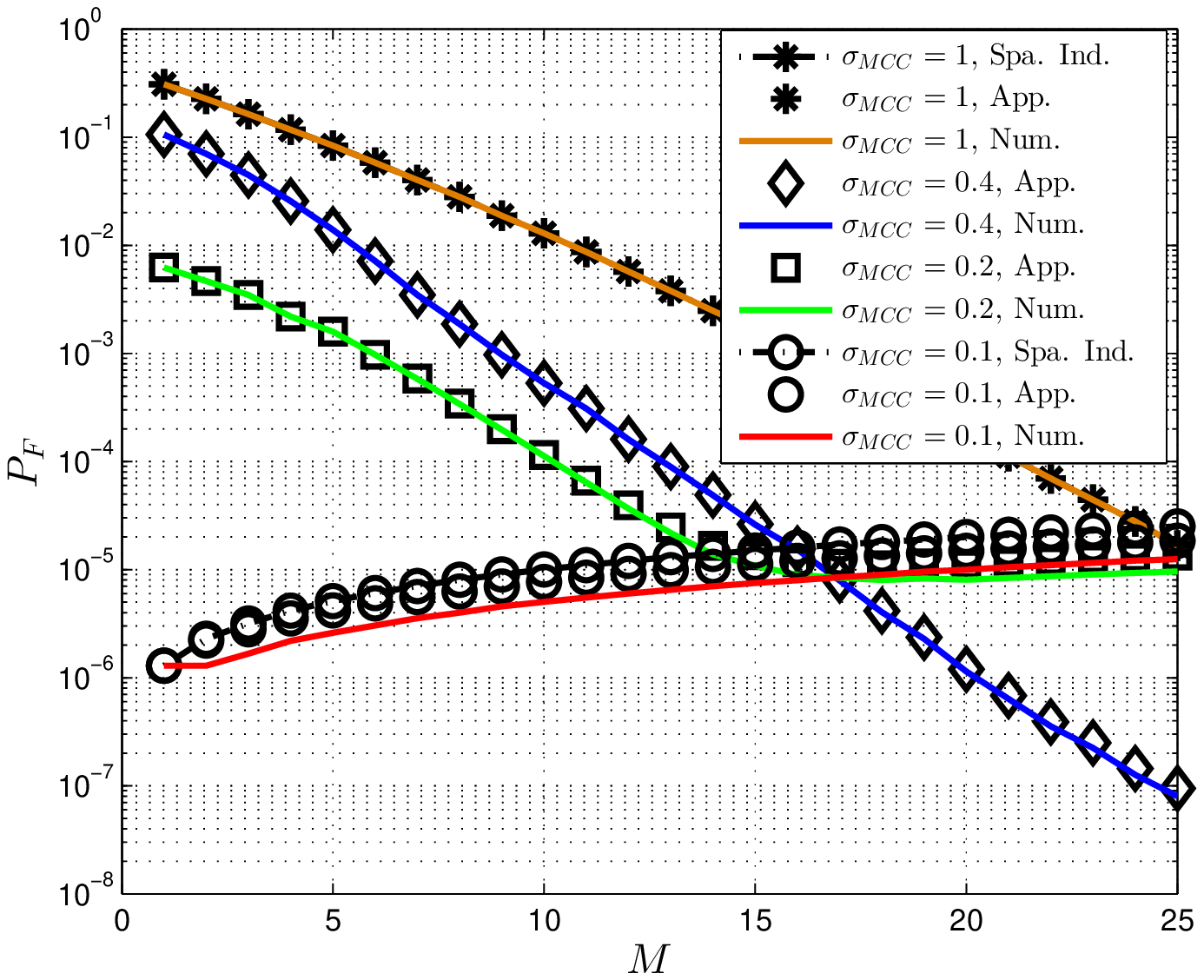}
                \label{fig:8}
        \end{subfigure}%
        \begin{subfigure}[b]{0.49\textwidth}
                \caption{}
                \includegraphics[width=\textwidth]{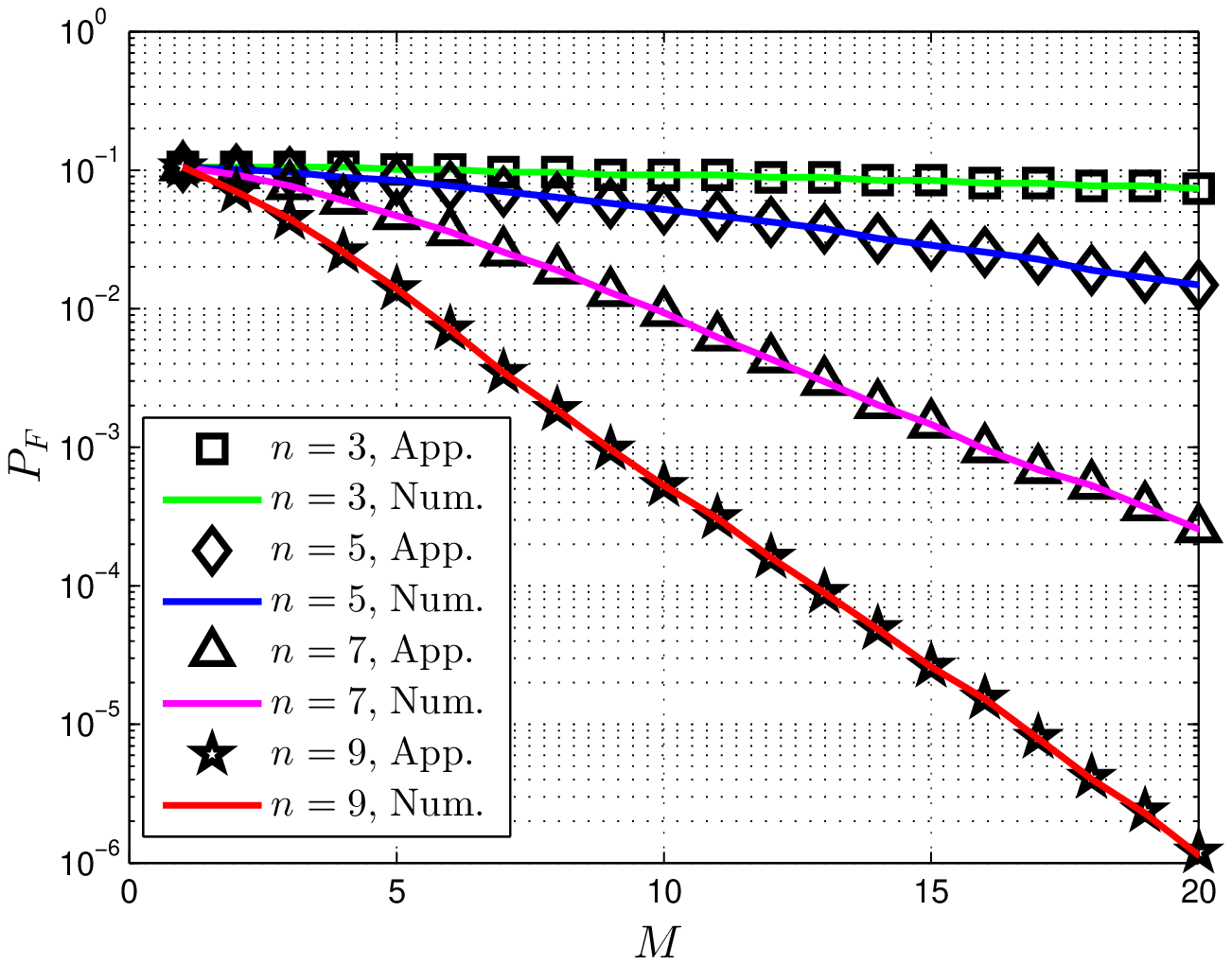}
                \label{fig:9}
        \end{subfigure} \vspace{-1em}
        \vspace{-1.5\baselineskip}
        \caption {(a) $P_F$ in terms of $M$ for different values of $\sigma_\mathit{MCC}$, in spatially  correlated/independent and temporary correlated scenario with $\rho  = 0.2$. (b) $P_F$ in terms of $M$ for different values of $n$ in spatially and temporally correlated scenario, for $\rho  = 0.2$ and ${\sigma _{\mathit{MCC}}} = 0.4$.} \label{fig:P_F}
\end{figure}

\begin{figure}[t!]
        \centering
        \begin{subfigure}[b]{0.49\textwidth}
                \caption{}
                \includegraphics[width=\textwidth]{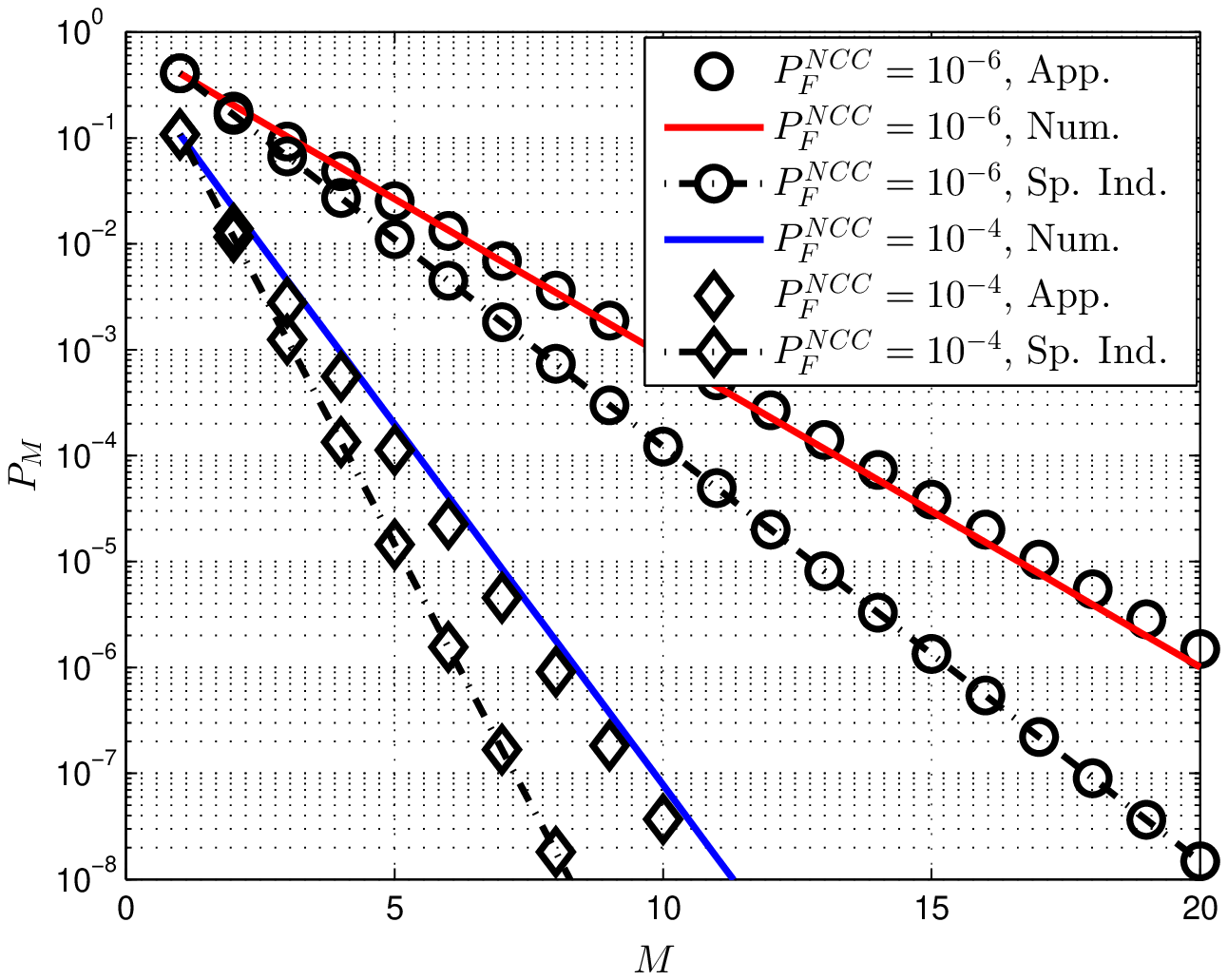}
                \label{fig:10}
        \end{subfigure}%
        \begin{subfigure}[b]{0.49\textwidth}
                \caption{}
                \includegraphics[width=\textwidth]{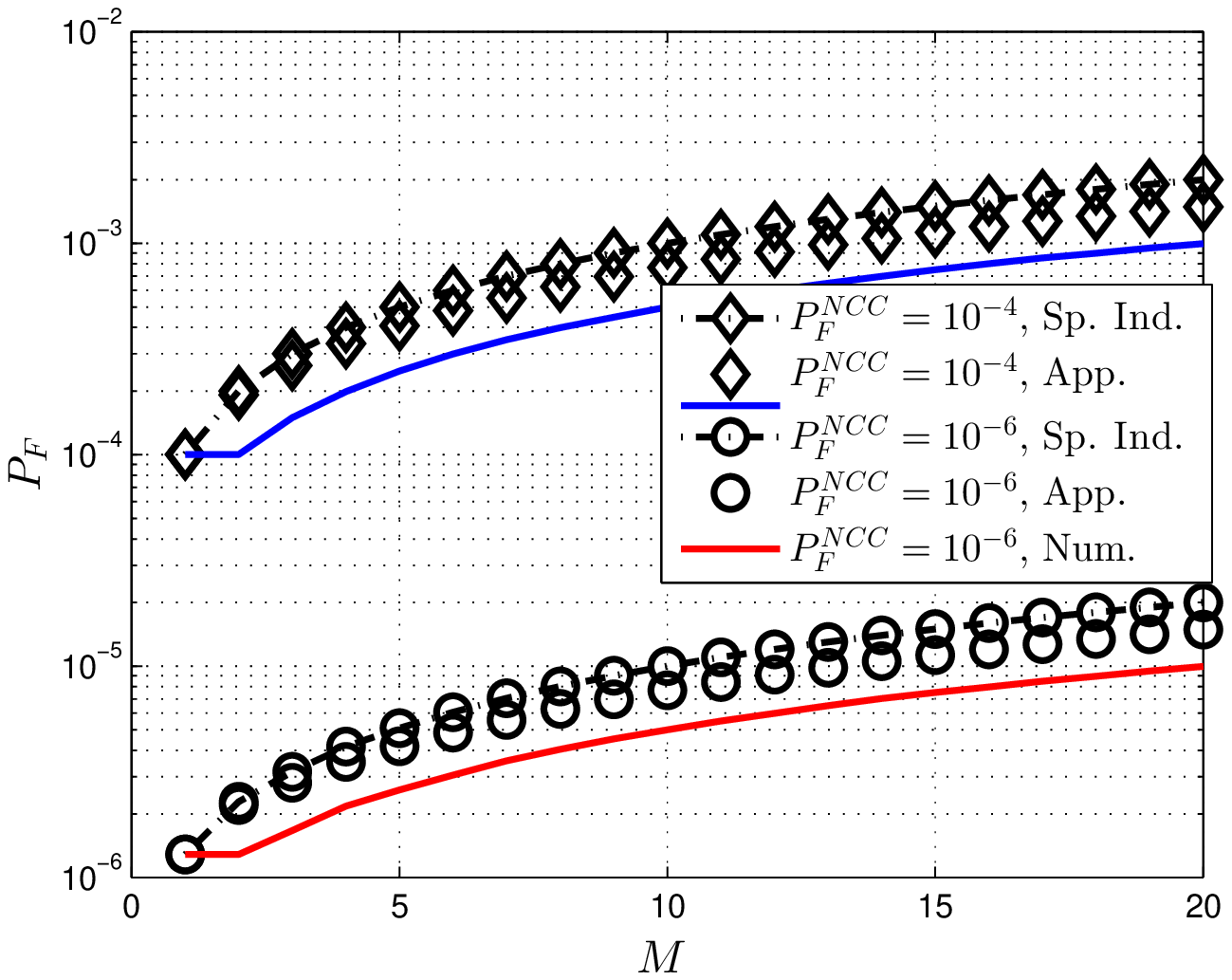}
                \label{fig:11}
        \end{subfigure}
        \vspace{-1.5\baselineskip}\\
        \begin{subfigure}[b]{0.49\textwidth}
                \caption{}
                \includegraphics[width=\textwidth]{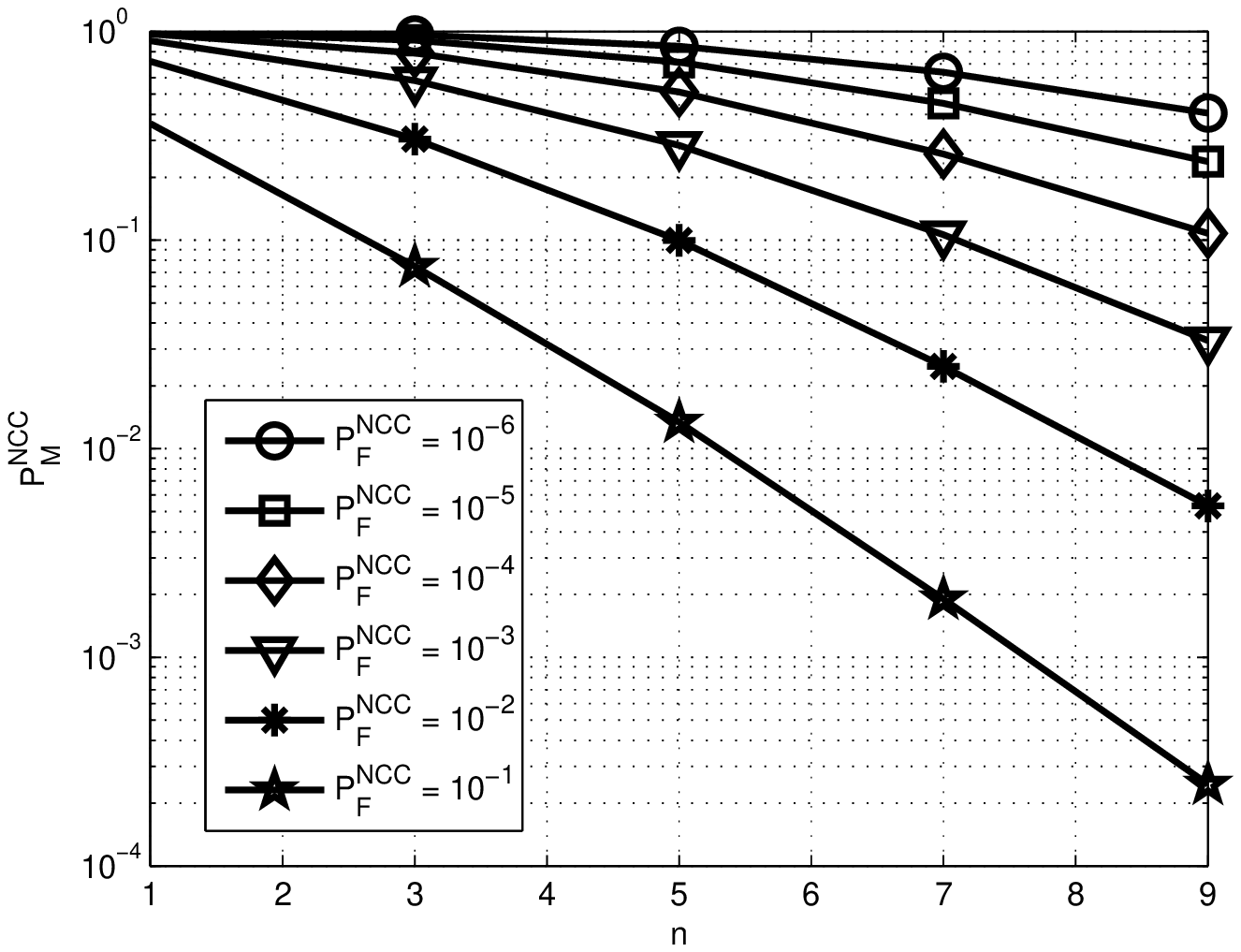}
                \label{fig:12}
        \end{subfigure}
        \begin{subfigure}[b]{0.49\textwidth}
                \caption{}
                \includegraphics[width=\textwidth]{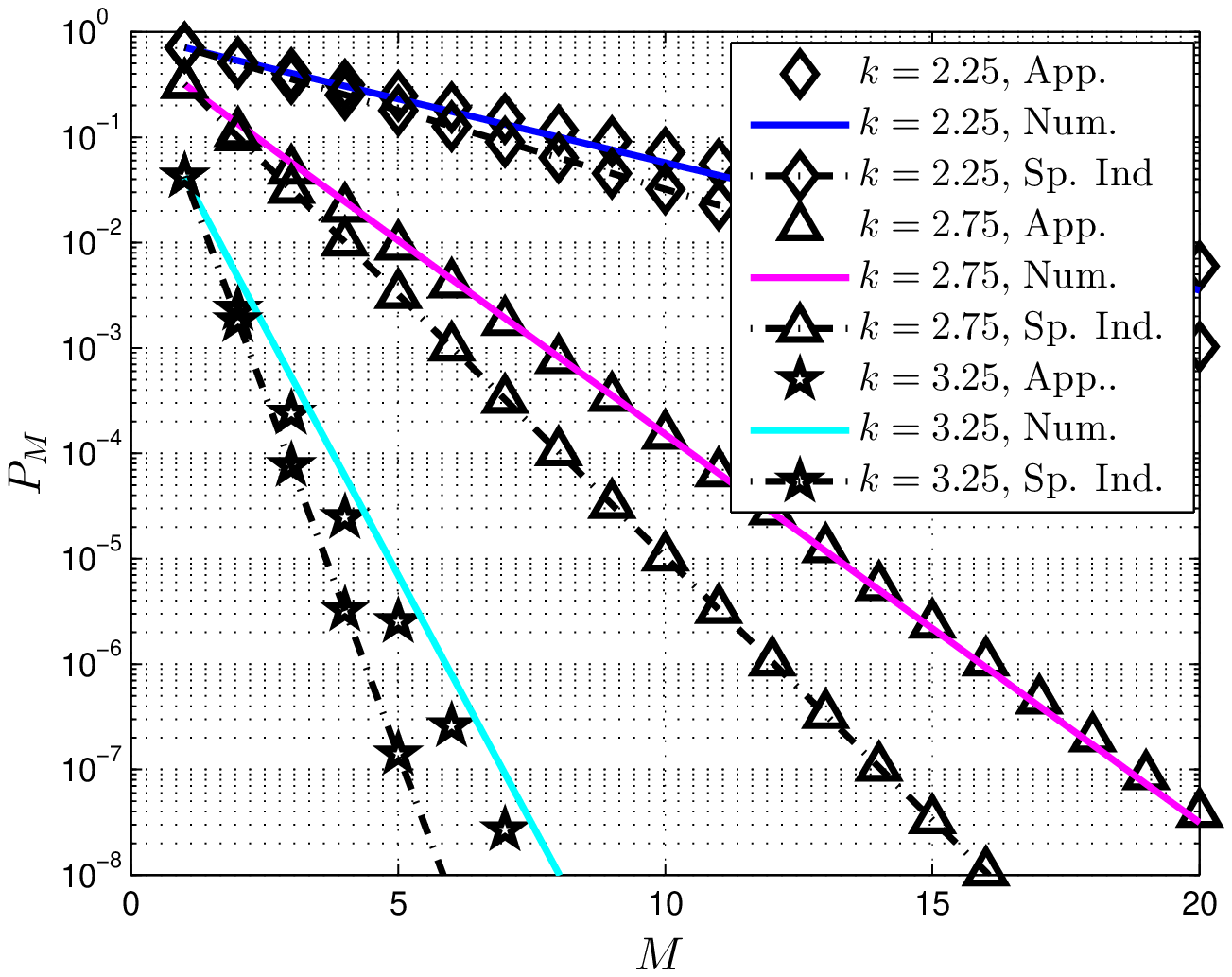}
                \label{fig:13}
        \end{subfigure}

        \vspace{-1.5\baselineskip}
        \caption {(a) $P_M$ in terms of $M$ for different values of $P_F^\mathit{NCC}$. (b) $P_F$  in terms of $M$ for different values of $P_F^\mathit{NCC}$  all for $\rho  = 0.2$ and ${\sigma _{\mathit{MCC}}} = 0.1$. (c) $P_M^{NCC}$ in terms of $n$ for different values of $P_F^\mathit{NCC}$, $\rho  = 0.2.$ (d) $P_M$ in terms of $M$ for different values of $k$, $\rho  = 0.2$ and ${\sigma _{\mathit{MCC}}} = 0.1$.} \label{fig:P_FNCC}
\end{figure}

Table~\ref{Table4} explains the presentation of numerical results in Figs~\ref{fig:4}-~\ref{fig:13}. Two methods for obtaining the performance results are considered, which are labeled as approximate and numerical in the sequel. First, we elaborate the results and comment on how the two methods are compared. Fig.~\ref{fig:4} shows the probability of miss-detection, ${P_M}$ in terms of the number of SNMs in the sample size, $M$ for different values of observation time, $n$ and temporal correlation $\rho$ in spatially independent scenario. It is evident that even a small value of temporal correlation, e.g., $\rho  = 0.1$, greatly affects $P_M$ . Fig.~\ref{fig:5} shows $P_M$ in terms of $M$, for different values of observation time, $n$ in spatially correlated and temporally independent scenario. One sees that spatial correlation of SNM observations degrades ${P_M}$. For example, with $M = 8$ and $n = 9$, spatially independent SNM observations results in a 20 times smaller ${P_M}$ when compared to the spatially correlated setting. Hence, if observations of SNMs are spatially correlated and we consider them as spatially independent, the reliability of NADS is substantially degraded. Figs~\ref{fig:6} and ~\ref{fig:7} show ${P_M}$ with ${\sigma _{\mathit{MCC}}} = 0.1$ and ${\sigma _{\mathit{MCC}}} = 0.4$ in terms of $M$ for different values of $n$ and $\rho$ in spatially and temporally correlated scenario. One sees that increasing ${\sigma_{\mathit{MCC}}}$ degrades $P_M$. Based on results in Figs~\ref{fig:4}-~\ref{fig:7}, it is evident that the probability of miss-detection $P_M$ obtained by the approximate method matches well with that computed based on the numerical method. Hence, the approximate method can be efficiently used to solve the design problem of~\eqref{eq:15}.

Fig.~\ref{fig:8} shows the probability of false alarm $P_F$ in terms of $M$, for different values of ${\sigma _{\mathit{MCC}}}$. One sees that the behavior of ${P_F}$  in terms of $M$ varies as ${\sigma _{\mathit{MCC}}}$ increases. For small values of ${\sigma _{\mathit{MCC}}}$, the performance degradation is due to error in the NCC. As evident in~\eqref{eq:48}, this performance result is valid for all values of $n$ and $\rho $, since in this experiment, $P_{F}^\mathit{NCC}$ is small (set to equality in~\eqref{eq:66}) and $\sigma_{\mathit{MCC}}$ is also small (effect of MCC is negligible on $P_F$). Moreover, one sees that any spatial or temporal correlation in SNM observations improves $P_F$. Fig.~\ref{fig:9} shows $P_F$ in terms of $M$ for different values of $n$ with $\sigma_{\mathit{MCC}} = 0.4$. One sees that the point at which the behavior of the curves changes depends both on ${\sigma _{\mathit{MCC}}}$ and $n$. From Figs~\ref{fig:8} and ~\ref{fig:9}, it is evident that the approximate method for computing $P_F$ slightly overestimates the false alarm probability when compared to the numerical method (see Table~\ref{Table4} for all values of $M$ and $n$. Hence, to avoid calculating multiple integrals in~\eqref{eq:35} and ~\eqref{eq:37}, the proposed approximate method may be efficiently used to address the design problem in~\eqref{eq:15}.

 The presented results in recent figures may be used to solve the optimization problem of~\eqref{eq:15}. For example, Figs~\ref{fig:5} and~\ref{fig:8} reveal that the optimized number of SNMs per unit size is $M = 7$ for ${\sigma _{\mathit{MCC}}} = 0.1$, when NCCs are spatially correlated and temporary independent for $\xi  = 1 - {10^{ - 6}}$, $\gamma  = {10^{ - 5}}$, $\sigma _{\mathit{MCC}}^{} = 0.1$ and $n = 9$. Figs~\ref{fig:7} and~\ref{fig:9} indicate that in the same setting and temporary and spatially correlated NCCs with ${\sigma _{\mathit{MCC}}} = 0.4$, we need to select $M = 23$. Furthermore, Fig.~\ref{fig:5} demonstrates a smaller ${P_M}$ when SNM observations are spatially independent as opposed to when they are correlated. Hence, if we consider the correlated observations as independent observations in the analyses instead, we will underestimate the required number of SNMs, $M$. For example, in the same setting with $P_F^{} = {10^{ - 5}}$ to achieve ${P_M} = {10^{ - 6}}$ we find $M = 10$ for spatially independent SNM observations. However, in the correlated scenario, we need at least $M = 13$.

Figs~\ref{fig:10} and ~\ref{fig:11} show $P_M$ and $P_F$ for NADS in terms of $M$, for different values of $P_F^{\mathit{NCC}}$ for spatially and temporary correlated NCCs. In Fig.~\ref{fig:11}, the results demonstrate that ${P_F}$ increases with $P_F^{\mathit{NCC}}$ and $M$. The typical trade-off of false alarm and detection performance of SNM over the NCC is visible in Fig.~\ref{fig:12}. Interestingly, $P_F^{\mathit{NCC}}$ affects the overall detection performance of NADS in the same way (Fig.~\ref{fig:10}), as it directly influences the NCC detection performance $P_M^{\mathit{NCC}}$(Fig.~\ref{fig:12}). These figures also demonstrate the effect of networking of the SNMs on the performance. Consider the performance of a single SNM in Fig.~\ref{fig:12} at $P_F^{\mathit{NCC}} = {10^{ - 6}}$ and $P_M^{\mathit{NCC}} \approx 0.35$. According to results in ~\ref{fig:10} and ~\ref{fig:11}, utilizing 20 SNMs leads to significantly improved $P_M$  of $10^{-6}$ and $P_F^{} \approx {10^{ - 5}}$ .

Our experiments (not reported here) reveal that the probability of miss-detection over the NCC noticeably reduces as parameter $k$ increases (this parameter may be used to indicate the disease progress level). Such a behavior then reflects in the overall system performance as depicted in Fig.~\ref{fig:13}. One sees that as $k$ increases, ${P_M}$ reduces much faster with $M$. The results indicate that if the competitor cell affects the molecular environment more strongly, the proposed NADS detects its presence more easily. A larger value of $k$ in~\eqref{eq:14}, may be interpreted as a disease which has progressed further and hence has altered the status of the molecular environment more significantly from a healthy setting.

The setting of this paper in the special case of spatial and temporal independent noise of SNMs reduces to that of our earlier study in~\cite{21}. However, the presented analysis in this work is exact in the said setting, whereas the prior work relies on certain approximations. Specifically, our extensive numerical results show that the approximated analyses of PF and PM in~\cite{21} are respectively a good approximation and an upper bound for the exact values presented here.

\begin{table}
    \centering
    \caption{Parameters of numerical results for Figs~\ref{fig:3}-\ref{fig:13}, X:Y:Z denotes the range of parameter as [X,Z] with step size Y, $G = 1$,
    $\mathit{NH} = 1$  , $vol = 1000 [nm^3]$. }
    \label{Table3}
    \small
    \begin{tabular}{| l | l | l | l | l |}
    \hline
    Parameter & ${\sigma _{\mathit{MCC}}}$ & ${\eta _1}$ & $n$ & $k$ \\ \hline
    Figs~\ref{fig:4}-\ref{fig:6} & 0.1 & $10^{-6}$ & 1:2:9& 2 \\ \hline
    Figs~\ref{fig:7} and \ref{fig:9} & 0.4 & $10^{-6}$ & 1:2:9 & 2 \\ \hline
    Fig.~\ref{fig:8} & 0.1, 0.2:0.2:1 & $10^{-6}$ & 9 & 2 \\ \hline
    Figs~\ref{fig:10}-~\ref{fig:12} & 0.1 & $[10^{-6} 10^{-5} 10^{-4} 10^{-3} 10^{-2} 10^{-1}]$& 9 & 2 \\ \hline
    Fig.~\ref{fig:13} & 0.1 & 0.1 & 1 & 1.75:0.5:3.25\\ \hline
    \end{tabular}
\end{table}

\begin{table}
    \centering
    \caption{Numerical results presentation specification for Figs~\ref{fig:P_M}-\ref{fig:P_FNCC}}
    \label{Table4}
    \small
    \begin{tabular}{| p{3cm} | p{2cm} | p{2cm} | p{2.5cm} | p{2.5cm} | p{2cm}|}
    \hline
    Method & Type of Curve & Spatially correlated & $Q_D$,$Q_F$ & $p'_l$,$p''_l$ & $P_M$, $P_F$\\ \hline
    Numerical (Num.) & Solid lines & $\surd$ & Numerically computed by~\eqref{eq:35} and ~\eqref{eq:37}& Approximated by $p'_l$  and $p''_l$  in~\eqref{eq:51} and ~\eqref{eq:52} & \eqref{eq:43} and \eqref{eq:44}
 \\ \hline
    Approximated (App.) & Markers only& $\surd$ & Approximated by ${\tilde Q_D}$ and ${\tilde Q_F}$ ~\eqref{eq:54} and ~\eqref{eq:55}& Approximated by $p'_l$  and $p''_l$  in~\eqref{eq:51} and ~\eqref{eq:52} & \eqref{eq:43} and \eqref{eq:44}
 \\ \hline
    Exact formula spatially independent scenario (Sp. Ind.) & Dashed-dotted lines & $\times$ & ~\eqref{eq:38} and ~\eqref{eq:39}& ~\eqref{eq:49} and ~\eqref{eq:50} & \eqref{eq:47} and \eqref{eq:48} \\ \hline
    \end{tabular}
\end{table}
 \vspace{-0.5\baselineskip}

%

\section{Conclusions}
An abnormality detection scheme for detection of competitor cells in a bio-molecular nano-network was proposed. This is motivated for the early detection and classification of diseases and enabling their timely and effective treatment. The proposed NADS is a two-tier network. The sensor nano-machines at the first tier act as receivers of a nano-communications channel modeling the molecular environment. The SNMs then communicate over a noisy channel to a data gathering node, which operates based on an OR fusion rule. The average number of received molecules serves as a feature for detecting the abnormalities at the SNMs. The detection performance of each SNM in presence of Gaussian observation noise was analyzed using a generalized likelihood ratio test. Moreover, the effects of temporal and spatial correlations of the SNMs observations on the detection performance were studied. The reported experiments results reveal that otherwise ignoring possibly existing temporal or spatial correlations would lead to noticeably inaccurate performance results. Next, quantifying the overall NADS detection performance, a design problem was set up that quantifies the minimum required concentration of SNMs for a desired level of NADS reliability. The solution determines the optimized operation of detectors for each of the NADS tiers. This in turn facilitates optimized abnormality detection with smallest possible side effects due to the injection of nano-sensors. The results indicate how effective fusion of the noisy observations collected from a number of sensor nano-machines with limited capabilities could provide an acceptable detection performance.

At the current stage of research on detection of diseases at the nano-scale, there are still many interesting open research problems. Here, we state a few of them. In this paper, the detection feature is set based on a mathematical modeling and certain valid approximations. Developing more precise models or obtaining the exact detection feature based on experimental measurements in the target tissue is an interesting research avenue. The side effects of injected SNMs on the molecular environment play an important role in the accuracy of the model and the performance of NADS. Hence, studying those effects is another key aspect of research in this field. Designing practical SNMs for detection of cancer or other diseases and taking the experimental constraints of those SNMs into consideration within the proposed NADS framework poses a number of other interesting and important research problems. Medical imaging is one approach to detection over the MCC; other approaches includes ultrasonic or terahertz communications. Realistic modeling of MCC noise is an interesting issue for enhancing the NADS performance for abnormality detection in a biomolecular environment.

\section{Appendices}
\vspace{-0.5\baselineskip}
\subsection{Proof of Theorem 1}
\vspace{-0.5\baselineskip}
The GLRT~\cite{56} for hypothesis test of~\eqref{eq:16} is given by
\begin{equation}\label{eq:56}
\begin{array}{l}
\frac{{\mathop {\max }\limits_{{\mathit{NR}} \ne {\mathit{NH}}} P\left( {\left. {{{\bf{y}}_j^n}} \right|{\mathit{NR}}} \right)}}{{P\left( {\left. {{{\bf{y}}_j^n}} \right|{\mathit{NH}}} \right)}} = \frac{{P\left( {\left. {{{\bf{y}}_j^n}} \right|\widehat \mathit{NR}_j} \right)}}{{P\left( {\left. {{{\bf{y}}_j^n}} \right|{\mathit{NH}}} \right)}} = \\
\frac{{\frac{1}{{{{\left( {2\pi } \right)}^{{n \mathord{\left/
 {\vphantom {n 2}} \right.
 \kern-\nulldelimiterspace} 2}}}\sigma _{\mathit{NCC}}^n{{\left| {{\Omega ^T}} \right|}^{{1 \mathord{\left/
 {\vphantom {1 2}} \right.
 \kern-\nulldelimiterspace} 2}}}}}\exp \left( { - \frac{1}{{2\sigma _{\mathit{NCC}}^2}}{{\left( {{{\bf{y}}_j^n} - {\widehat {{\bf{NR}}}^n}} \right)}^\dag }{\Omega ^{{T^{ - 1}}}}\left( {{{\bf{y}}_j^n} - {\widehat {{\bf{NR}}}^n}} \right)} \right)}}{{\frac{1}{{{{\left( {2\pi } \right)}^{{n \mathord{\left/
 {\vphantom {n 2}} \right.
 \kern-\nulldelimiterspace} 2}}}\sigma _{\mathit{NCC}}^n{{\left| {{\Omega ^T}} \right|}^{{1 \mathord{\left/
 {\vphantom {1 2}} \right.
 \kern-\nulldelimiterspace} 2}}}}}\exp \left( { - \frac{1}{{2\sigma _{\mathit{NCC}}^2}}{{\left( {{{\bf{y}}_j^n} - {\bf{NH}}^n} \right)}^\dag }{\Omega ^{{T^{^{ - 1}}}}}\left( {{{\bf{y}}_j^n} - {\bf{NH}}^n} \right)} \right)}} > \tau,
\end{array}
\end{equation}
 where ${\widehat {{\bf{NR}}}^n} = {\left[ {\widehat {\mathit{NR}}_j} \right]_{1 \times n}}$, ${\bf{NH}}^n = {\left[ {\mathit{NH}} \right]_{1 \times n}}$ and ${\widehat {\mathit{NR}}_j}$ is given by~\eqref{eq:24}. Simplifying~\eqref{eq:56} we have
 \begin{equation}\label{eq:57}
 \exp \left\{ {\frac{{ - 1}}{{2\sigma _{NCC}^2}}\left[ {\sum\limits_{l = 1}^n {\sum\limits_{i = 1}^n {\left( {{y_{lj}} - \widehat \mathit{{NR}}_j} \right)} } \left( {{y_{ij}} - \widehat \mathit{{NR}}_j} \right)\psi _{il}^{TC} - \sum\limits_{l = 1}^n {\sum\limits_{i = 1}^n {\left( {{y_{lj}} - NH} \right)} } \left( {{y_{ij}} - NH} \right)\psi _{il}^{TC}} \right]} \right\} > \tau .
\end{equation}
Computing the natural Logarithm of~\eqref{eq:57}, we obtain
\begin{equation}\label{eq:58}
\begin{array}{l}
\frac{{ - 1}}{{2\sigma _{NCC}^2}}\left\{ {\sum\limits_{l = 1}^n {\sum\limits_{i = 1}^n {\left( {{y_{lj}} - \widehat \mathit{NR}_j} \right)} } \left( {{y_{ij}} - \widehat \mathit{NR}_j} \right)\psi _{il}^{TC} - } \right.\\
\left. {\sum\limits_{l = 1}^n {\sum\limits_{i = 1}^n {\left( {\left( {{y_{lj}} - \widehat \mathit{NR}_j} \right) + \left( {\widehat \mathit{NR}_j - {\mathit{NH}}} \right)} \right)} } \left( {\left( {{y_{ij}} - \widehat \mathit{NR}_j} \right) + \left( {\widehat \mathit{NR}_j - \mathit{NH}} \right)} \right)\psi _{il}^{TC}} \right\} > \log \tau .
\end{array}
\end{equation}
Following some manipulations, we have
\begin{equation}\label{eq:59}
\begin{array}{l}
\frac{{ - 1}}{{2\sigma _{\mathit{NCC}}^2}}\left\{ { - \sum\limits_{l = 1}^n {\sum\limits_{i = 1}^n {\left( {{y_{lj}} - \widehat \mathit{NR}_j} \right)\left( {\widehat \mathit{NR}_j - {\mathit{NH}}} \right)} } } \right.\psi _{il}^{\mathit{TC}} - \\
\left. {{\rm{ }}\sum\limits_{l = 1}^n {\sum\limits_{i = 1}^n {\left( {{y_{ij}} - \widehat \mathit{NR}_j} \right)\left( {\widehat \mathit{NR}_j - {\mathit{NH}}} \right)\psi _{il}^{\mathit{TC}}} }  - \sum\limits_{l = 1}^n {\sum\limits_{i = 1}^n {{{\left( {\widehat \mathit{NR}_j - {\mathit{NH}}} \right)}^2}\psi _{il}^{\mathit{TC}}} } } \right\} > \log \tau .
\end{array}
\end{equation}
Replacing $\widehat \mathit{NR}_j$ from~\eqref{eq:24} in the first and second terms of RHS of~\eqref{eq:59}, we have
\begin{equation}\label{eq:60}
\frac{{ - 1}}{{2\sigma _{\mathit{NCC}}^2}}\left[ { - \sum\limits_{l = 1}^n {\sum\limits_{i = 1}^n {{{\left( {\widehat \mathit{NR}_j - \mathit{NH}} \right)}^2}\psi _{il}^{\mathit{TC}}} } } \right] > \log \tau  \Rightarrow {\left( {\widehat \mathit{NR}_j - {\mathit{NH}}} \right)^2} > \frac{{2\sigma _{\mathit{NCC}}^2\log \tau }}{{\sum\limits_{l = 1}^n {\sum\limits_{i = 1}^n {\psi _{il}^{\mathit{TC}}} } }}.
\end{equation}
With more simplification, the decision region for hypothesis test of~\eqref{eq:16} is obtained as
\begin{equation}\label{eq:61}
\left\{ {\begin{array}{*{20}{c}}
{{H_0}}&{{\mathit{NH}} - \tau ' < \widehat \mathit{NR}_j < {\mathit{NH}} + \tau '}\\
{{H_1}}&{\begin{array}{*{20}{c}}
{\widehat \mathit{NR}_j > {\mathit{NH}} + \tau '}\\
{\widehat \mathit{NR}_j < {\mathit{NH}} - \tau ',}
\end{array}}
\end{array}} \right.
\end{equation}
where
\begin{equation}\label{eq:62}
\tau ' \buildrel \Delta \over = \sqrt {{{2\sigma _{\mathit{NCC}}^2\log \left( \tau  \right)} \mathord{\left/
 {\vphantom {{2\sigma _{\mathit{NCC}}^2\log \left( \tau  \right)} {\sum\limits_{l = 1}^n {\sum\limits_{i = 1}^n {\psi _{il}^{\mathit{TC}}} } }}} \right.
 \kern-\nulldelimiterspace} {\sum\limits_{l = 1}^n {\sum\limits_{i = 1}^n {\psi _{il}^{\mathit{TC}}} } }}}.
\end{equation}
For deriving the false-alarm probability of a decision rule, we need to calculate the PDF of $\widehat \mathit{NR}_j$ given in~\eqref{eq:24}. The random variables $y_{ij}$ are jointly Gaussian, hence their weighted summation is also Gaussian. When ${H_0}$ is true, the mean of decision variable at the SNM, $\widehat \mathit{NR}_j$ is $\mathit{NH}$ and its variance is given by~\cite{61}
\begin{equation}\label{eq:63}
\sigma _D^2 = {\left( {\sum\limits_{l = 1}^n {\sum\limits_{i = 1}^n {\psi _{il}^{\mathit{TC}}} } } \right)^{ - 2}}\left\{ {\sum\limits_{l = 1}^n {{\mathop{\rm var}} ({\lambda _l}) + 2} \sum\limits_{q < l}^n {{\mathop{\rm cov}} ({\lambda _q},{\lambda _l})} } \right\}.
\end{equation}
in which ${\lambda _l} = {y_{lj}}\sum\limits_{i = 1}^n {\psi _{il}^{\mathit{TC}}}$, the subscript of $j$ in $\lambda_l$ is dropped due to homogeneous assumption of NCCs for each SNM,~${\mathop{\rm var}} ({\lambda _l}) = {\left( {\sum\limits_{i = 1}^n {\psi _{il}^{\mathit{TC}}} } \right)^2}\sigma _{\mathit{NCC}}^2$,~and~${\mathop{\rm cov}} ({\lambda _q},{\lambda _l}) = \left( {\sum\limits_{i = 1}^n {\psi _{il}^{\mathit{TC}}} } \right)\left( {\sum\limits_{i = 1}^n {\psi _{il}^{\mathit{TC}}} } \right)\omega _{ql}^{\mathit{TC}}$ $\sigma _{\mathit{NCC}}^2$. We have
\begin{equation}\label{eq:64}
\sigma _D^2 = {\left( {\sum\limits_{l = 1}^n {\sum\limits_{i = 1}^n {\psi _{il}^{\mathit{TC}}} } } \right)^{ - 2}}\left\{ {\sum\limits_{l = 1}^n {{{\left( {\sum\limits_{i = 1}^n {\psi _{il}^{\mathit{TC}}} } \right)}^2}\left( {\sigma _{\mathit{NCC}}^2 + 2\sum\limits_{l = 1}^{n - 1} {\sum\limits_{q = l + 1}^n {\omega _{ql}^{\mathit{TC}}\sigma _{\mathit{NCC}}^2} } } \right)} } \right\}.
\end{equation}
Therefore, the false alarm probability can be expressed by
\begin{equation}\label{eq:65}
P_F^{\mathit{NCC}} = 1 - \Pr \left\{ {\left. {\mathit{NH} - \tau ' < {{\cal N}}\left( {\mathit{NH},\sigma _D^2} \right) < \mathit{NH} + \tau '} \right|{H_0}} \right\},
\end{equation}
and as we desire to have $P_F^{\mathit{NCC}} \le {\eta _1}$, we obtain
\begin{equation}\label{eq:66}
2\phi \left( {{{\tau '} \mathord{\left/
 {\vphantom {{\tau '} {\sigma _D^{}}}} \right.
 \kern-\nulldelimiterspace} {\sigma _D^{}}}} \right) - 1 \ge 1 - {\eta _1},
\end{equation}
that is satisfied with equality, when we have
\begin{equation}\label{eq:67}
\tau ' = \sigma _D^{}{\phi ^{ - 1}}\left( {1 - \frac{{{\eta _1}}}{2}} \right).
\end{equation}
According to the region of ${H_1}$ in~\eqref{eq:61}, $P_D^{\mathit{NCC}}$ can be written as
\begin{equation}\label{eq:68}
P_D^{\mathit{NCC}} = \Pr \left\{ {\left. {\widehat \mathit{NR}_j < {\mathit{NH}} - \tau ',\widehat \mathit{NR}_j > \mathit{NH} + \tau '} \right|{H_1}} \right\},
\end{equation}
and due to the Gaussian distribution of $\widehat \mathit{NR}_j$, we have
\begin{equation}\label{eq:69}
\begin{array}{l}
P_D^{\mathit{NCC}} = \Pr \left\{ {{{\cal N}}\left( {{\mathit{NR}},\sigma _D^2} \right) < {\mathit{NH}} - \tau ',{{\cal N}}\left( {{\mathit{NR}},\sigma _D^2} \right) > {\mathit{NH}} + \tau '} \right\} = \\
1 - {{\cal Q}}\left( {\frac{{{\mathit{NH}} - \tau ' - {\mathit{NR}}}}{{\sigma _D^2}}} \right) + {{\cal Q}}\left( {\frac{{{\mathit{NH}} + \tau ' - {\mathit{NR}}}}{{\sigma _D^2}}} \right).
\end{array}
\end{equation}
The relation of ${P_D}$ and $\mathit{NR}$ is evident in~\eqref{eq:69}. When a competitor cell is present in the NCC environment, $\mathit{NR}$ deviates from the $\mathit{NH}$. For a specific type of competitor cell, considering~\eqref{eq:14} in~\eqref{eq:69}, $P_D^{\mathit{NCC}}$  can be obtained as
\vspace{-0.5\baselineskip}
\begin{equation}\label{eq:70}
\begin{array}{l}
P_D^{\mathit{NCC}} = 1 - {{\cal Q}}\left( {{{\left( {{\mathit{NH}} - \tau ' - \left( {1 \pm k{\sigma _{\mathit{NCC}}}} \right){\mathit{NH}}} \right)} \mathord{\left/
 {\vphantom {{\left( {{\mathit{NH}} - \tau ' - \left( {1 \pm k{\sigma _{\mathit{NCC}}}} \right){\mathit{NH}}} \right)} {{\sigma _D}}}} \right.
 \kern-\nulldelimiterspace} {{\sigma _D}}}} \right) +{{\cal Q}}\left( {{{\left( {{\mathit{NH}} + \tau ' - \left( {1 \pm k{\sigma _{\mathit{NCC}}}} \right){\mathit{NH}}} \right)} \mathord{\left/
 {\vphantom {{\left( {{\mathit{NH}} + \tau ' - \left( {1 \pm k{\sigma _{\mathit{NCC}}}} \right){\mathit{NH}}} \right)} {{\sigma _D}}}} \right.
 \kern-\nulldelimiterspace} {{\sigma _D}}}} \right)\\
{\rm{       }} = 1 - {{\cal Q}}\left( {{{\left( { - \tau ' \mp k{\sigma _{\mathit{NCC}}}{\mathit{NH}}} \right)} \mathord{\left/
 {\vphantom {{\left( { - \tau ' \mp k{\sigma _{\mathit{NCC}}}{\mathit{NH}}} \right)} {\left( {{\sigma _D}} \right)}}} \right.
 \kern-\nulldelimiterspace} {\left( {{\sigma _D}} \right)}}} \right) + {{\cal Q}}\left( {{{\left( {\tau ' \mp k{\sigma _{\mathit{NCC}}}{\mathit{NH}}} \right)} \mathord{\left/
 {\vphantom {{\left( {\tau ' \mp k{\sigma _{\mathit{NCC}}}{\mathit{NH}}} \right)} {\left( {{\sigma _D}} \right)}}} \right.
 \kern-\nulldelimiterspace} {\left( {{\sigma _D}} \right)}}} \right).
\end{array}
\end{equation}
Replacing $\tau '$ from~\eqref{eq:67} in~\eqref{eq:70}, the following result is obtained
\begin{equation}\label{eq:71}
\begin{array}{l}
P_D^{\mathit{NCC}} = 1 - \\
{{\cal Q}}\left( {{{\left( { - {\sigma _D}{\phi ^{ - 1}}\left( {1 - \frac{{{\eta _1}}}{2}} \right) \mp k{\sigma _{\mathit{NCC}}}{\mathit{NH}}} \right)} \mathord{\left/
 {\vphantom {{\left( { - {\sigma _D}{\phi ^{ - 1}}\left( {1 - \frac{{{\eta _1}}}{2}} \right) \mp k{\sigma _{\mathit{NCC}}}{\mathit{NH}}} \right)} {{\sigma _D}}}} \right.
 \kern-\nulldelimiterspace} {{\sigma _D}}}} \right) + {{\cal Q}}\left( {{{\left( {{\sigma _D}{\phi ^{ - 1}}\left( {1 - \frac{{{\eta _1}}}{2}} \right) \mp k{\sigma _{\mathit{NCC}}}{\mathit{NH}}} \right)} \mathord{\left/
 {\vphantom {{\left( {{\sigma _D}{\phi ^{ - 1}}\left( {1 - \frac{{{\eta _1}}}{2}} \right) \mp k{\sigma _{\mathit{NCC}}}{\mathit{NH}}} \right)} {{\sigma _D}}}} \right.
 \kern-\nulldelimiterspace} {{\sigma _D}}}} \right).
\end{array}
\end{equation}

\subsection{Proof of Lemma 1}
It is evident in~\eqref{eq:24} that $\widehat \mathit{NR}_j$ is a weighted sum of $y_{ij}$'s, which are jointly Gaussian distributed and hence their summation is Gaussian with mean $\mathit{NR}$.
The correlation coefficient of $\widehat \mathit{NR}_j$  and $\widehat \mathit{NR}_q$  is given by
\vspace{-0.5\baselineskip}
\begin{equation}\label{eq:72}
{\rm{cor}}(\widehat \mathit{NR}_j,\widehat \mathit{NR}_q) = \frac{{E\left[ {\left( {\widehat \mathit{NR}_j - {\mathit{NR}}} \right)\left( {\widehat \mathit{NR}_q - \mathit{NR}} \right)} \right]}}{{\sqrt {E\left[ {{{\left( {\widehat \mathit{NR}_j - {\mathit{NR}}} \right)}^2}} \right]E\left[ {{{\left( {\widehat \mathit{NR}_q - {\mathit{NR}}} \right)}^2}} \right]} }}.
\end{equation}
\vspace{-0.5\baselineskip}
Using~\eqref{eq:24}, we obtain
\begin{equation}\label{eq:73}
\begin{array}{l}
{\rm{cor}}(\widehat \mathit{NR}_j,\widehat \mathit{NR}_q) = \\
\frac{{E\left[ {\left( {{{\sum\limits_{l = 1}^n {\sum\limits_{i = 1}^n {{y_{lj}}\psi _{il}^{\mathit{TC}}} } } \mathord{\left/
 {\vphantom {{\sum\limits_{l = 1}^n {\sum\limits_{i = 1}^n {{y_{lj}}\psi _{il}^{\mathit{TC}}} } } {\sum\limits_{l = 1}^n {\sum\limits_{i = 1}^n {\psi _{il}^{\mathit{TC}}} } }}} \right.
 \kern-\nulldelimiterspace} {\sum\limits_{l = 1}^n {\sum\limits_{i = 1}^n {\psi _{il}^{\mathit{TC}}} } }} - {\mathit{NR}}} \right)\left( {{{\sum\limits_{l = 1}^n {\sum\limits_{i = 1}^n {{y_{lq}}\psi _{il}^{\mathit{TC}}} } } \mathord{\left/
 {\vphantom {{\sum\limits_{l = 1}^n {\sum\limits_{i = 1}^n {{y_{lq}}\psi _{il}^{\mathit{TC}}} } } {\sum\limits_{l = 1}^n {\sum\limits_{i = 1}^n {\psi _{il}^{\mathit{TC}}} } }}} \right.
 \kern-\nulldelimiterspace} {\sum\limits_{l = 1}^n {\sum\limits_{i = 1}^n {\psi _{il}^{\mathit{TC}}} } }} - {\mathit{NR}}} \right)} \right]}}{{\sqrt {E\left[ {{{\left( {{{\sum\limits_{l = 1}^n {\sum\limits_{i = 1}^n {{y_{lj}}\psi _{il}^{\mathit{TC}}} } } \mathord{\left/
 {\vphantom {{\sum\limits_{l = 1}^n {\sum\limits_{i = 1}^n {{y_{lj}}\psi _{il}^{\mathit{TC}}} } } {\sum\limits_{l = 1}^n {\sum\limits_{i = 1}^n {\psi _{il}^{\mathit{TC}}} } }}} \right.
 \kern-\nulldelimiterspace} {\sum\limits_{l = 1}^n {\sum\limits_{i = 1}^n {\psi _{il}^{\mathit{TC}}} } }} - {\mathit{NR}}} \right)}^2}} \right]E\left[ {{{\left( {{{\sum\limits_{l = 1}^n {\sum\limits_{i = 1}^n {{y_{lq}}\psi _{il}^{\mathit{TC}}} } } \mathord{\left/
 {\vphantom {{\sum\limits_{l = 1}^n {\sum\limits_{i = 1}^n {{y_{lq}}\psi _{il}^{\mathit{TC}}} } } {\sum\limits_{l = 1}^n {\sum\limits_{i = 1}^n {\psi _{il}^{\mathit{TC}}} } }}} \right.
 \kern-\nulldelimiterspace} {\sum\limits_{l = 1}^n {\sum\limits_{i = 1}^n {\psi _{il}^{\mathit{TC}}} } }} - {\mathit{NR}}} \right)}^2}} \right]} }},
\end{array}
\end{equation}
and following some mathematical manipulation, we have
\begin{equation}\label{eq:74}
{\rm{cor}}(\widehat \mathit{NR}_j,\widehat \mathit{NR}_q) = \frac{{E\left[ {\left( {\sum\limits_{l = 1}^n {\sum\limits_{i = 1}^n {\left( {{y_{lj}} - {\mathit{NR}}} \right)\psi _{il}^{\mathit{TC}}} } } \right)\left( {\sum\limits_{l = 1}^n {\sum\limits_{i = 1}^n {\left( {{y_{lq}} - {\mathit{NR}}} \right)\psi _{il}^{\mathit{TC}}} } } \right)} \right]}}{{\sqrt {E\left[ {{{\left( {\sum\limits_{l = 1}^n {\sum\limits_{i = 1}^n {\left( {{y_{lj}} - {\mathit{NR}}} \right)\psi _{il}^{\mathit{TC}}} } } \right)}^2}} \right]E\left[ {{{\left( {\sum\limits_{l = 1}^n {\sum\limits_{i = 1}^n {\left( {{y_{lq}} - {\mathit{NR}}} \right)\psi _{il}^{\mathit{TC}}} } } \right)}^2}} \right]} }}.
\end{equation}
Using the assumption of separability of spatial and temporal correlation as discussed in Section II and some mathematical manipulations, we have
\vspace{-0.5\baselineskip}
\begin{equation}\label{eq:75}
{\rm{cor}}(\widehat \mathit{NR}_j,\widehat \mathit{NR}_q) = \frac{{\omega _{jq}^{\mathit{SC}}\left( {\sum\limits_{l = 1}^n {\left( {\sigma _{\mathit{NCC}}^2\left( {\sum\limits_{i = 1}^n {\psi _{il}^{\mathit{TC}}\sum\limits_{i = 1}^n {\psi _{il}^{\mathit{TC}}} } } \right) + 2\sum\limits_{k = l + 1}^n {\omega _{kl}^{\mathit{TC}}\sigma _{\mathit{NCC}}^2\sum\limits_{i = 1}^n {\psi _{il}^{\mathit{TC}}\sum\limits_{i = 1}^n {\psi _{il}^{\mathit{TC}}} } } } \right)} } \right)}}{{\sum\limits_{l = 1}^n {\left\{ {{{\left( {\sum\limits_{i = 1}^n {\psi _{il}^{\mathit{TC}}} } \right)}^2}\sigma _{\mathit{NCC}}^2} \right.} {\rm{  + 2}}\sum\limits_{k = l + 1}^n {\left( {\sum\limits_{i = 1}^n {\psi _{il}^{\mathit{TC}}} \sum\limits_{i = 1}^n {\psi _{il}^{\mathit{TC}}} } \right)\left. {\omega _{kl}^{\mathit{TC}}\sigma _{\mathit{NCC}}^2} \right\}{\rm{ }}} }} = \omega _{jq}^{\mathit{SC}}.
\end{equation}
\vspace{-2.5\baselineskip}
\subsection{Proof of Theorem 2}
\vspace{-0.5\baselineskip}
The MAP rule is written as
\vspace{-0.5\baselineskip}
\begin{equation}\label{eq:76}
\begin{array}{l}
\mathop {\max }\limits_{i \in \left\{ {0,1} \right\}} P\left( {\left. {{H_i}} \right|V} \right)\mathop  = \limits^{({\rm{a}})} {\mathop{P}\nolimits} \left( {\left. {{H_i}} \right|V,{W_0}} \right)P\left( {\left. {{W_0}} \right|V} \right) + {\mathop{P}\nolimits} \left( {\left. {{H_i}} \right|V,{W_1}} \right)P\left( {\left. {{W_1}} \right|V} \right)\\
\mathop  = \limits^{(b)} \mathop {\max }\limits_{i \in \left\{ {0,1} \right\}} {\mathop{P}\nolimits} \left( {\left. {{H_i}} \right|{W_0}} \right)P\left( {\left. {{W_0}} \right|V} \right) + {\mathop{P}\nolimits} \left( {\left. {{H_i}} \right|{W_1}} \right)P\left( {\left. {{W_1}} \right|V} \right)\\
\mathop  = \limits^{(c)} \mathop {\max }\limits_{i \in \left\{ {0,1} \right\}} \frac{{{\mathop{P}\nolimits} \left( {\left. {{W_0}} \right|{H_i}} \right)P({H_i})}}{{P({W_0})}}P\left( {\left. {{W_0}} \right|V} \right) + \frac{{{\mathop{P}\nolimits} \left( {\left. {{W_1}} \right|{H_i}} \right)P({H_i})}}{{P({W_1})}}P\left( {\left. {{W_1}} \right|V} \right)
\end{array}
\end{equation}
where in the probability function of $P\left(  \bullet  \right)$ event of $V = v$ is denoted by $V$ for brevity and events of ${W_0}$/${W_1}$ are defined as
\begin{equation}\label{eq:77}
\left\{ {\begin{array}{*{20}{c}}
{{W_0}:\sum\nolimits_{j = 1}^M {{X_j}}  < G}\\
{{W_1}:\sum\nolimits_{j = 1}^M {{X_j}}  \ge G.}
\end{array}} \right.
\end{equation}
 In~\eqref{eq:76}, (a) follows from the law of total probability, (b) from Markov property of ${H_i} \to {W_i} \to V$ and (c) follows from the Bayes’ rule. Simplifying ~\eqref{eq:76}, we have
\begin{equation}\label{eq:78}
\begin{array}{l}
\frac{{{\mathop{\rm P}\nolimits} \left( {\left. {{W_0}} \right|{H_0}} \right)P({H_0})}}{{P({W_0})}}P\left( {\left. {{W_0}} \right|V} \right) + \frac{{{\mathop{\rm P}\nolimits} \left( {\left. {{W_1}} \right|{H_0}} \right)P({H_0})}}{{P({W_1})}}P\left( {\left. {{W_1}} \right|V} \right)\mathop {\mathop  > \limits^{{H_0}} }\limits_ <  \\
\frac{{{\mathop{\rm P}\nolimits} \left( {\left. {{W_0}} \right|{H_1}} \right)P({H_1})}}{{P({W_0})}}P\left( {\left. {{W_0}} \right|V} \right) + \frac{{{\mathop{\rm P}\nolimits} \left( {\left. {{W_1}} \right|{H_1}} \right)P({H_1})}}{{P({W_1})}}P\left( {\left. {{W_1}} \right|V} \right)\mathop  \Rightarrow \limits^{({\rm{a)}}} \\
\frac{{\left( {1 - {Q_F}} \right)P({H_0})}}{{P({W_0})}}P\left( {\left. {{W_0}} \right|V} \right) + \frac{{{Q_F}P({H_0})}}{{1 - P({W_0})}}P\left( {\left. {{W_1}} \right|V} \right)\mathop {\mathop  > \limits^{{H_0}} }\limits_ <  \frac{{\left( {1 - {Q_D}} \right)P({H_1})}}{{P({W_0})}}P\left( {\left. {{W_0}} \right|V} \right) + \frac{{{Q_D}P({H_1})}}{{1 - P({W_0})}}P\left( {\left. {{W_1}} \right|V} \right)\mathop  \Rightarrow \limits^{(b)} \\
\frac{{\left( {1 - {Q_F}} \right)P({H_0})}}{{P({W_0})}}\frac{{P\left( {\left. V \right|{W_0}} \right)P\left( {{W_0}} \right)}}{{P(V)}} + \frac{{{Q_F}P({H_0})}}{{1 - P({W_0})}}\frac{{P\left( {\left. V \right|{W_1}} \right)P\left( {{W_1}} \right)}}{{P(V)}}\mathop {\mathop  > \limits^{{H_0}} }\limits_ <  \\
\frac{{\left( {1 - {Q_D}} \right)P({H_1})}}{{P({W_0})}}\frac{{P\left( {\left. V \right|{W_0}} \right)P\left( {{W_0}} \right)}}{{P(V)}} + \frac{{{Q_D}P({H_1})}}{{1 - P({W_0})}}\frac{{P\left( {\left. V \right|{W_1}} \right)P\left( {{W_1}} \right)}}{{P(V)}} \Rightarrow \\
\left( {1 - {Q_F}} \right)P({H_0})P\left( {\left. V \right|{W_0}} \right) + {Q_F}P({H_0})P\left( {\left. V \right|{W_1}} \right)\mathop {\mathop  > \limits^{{H_0}} }\limits_ <  \left( {1 - {Q_D}} \right)P({H_1})P\left( {\left. V \right|{W_0}} \right) + {Q_D}P({H_1})P\left( {\left. V \right|{W_1}} \right)
\end{array}
\end{equation}
where, (a) is derived based on definition of ${Q_F}$ and ${Q_D}$ respectively in~\eqref{eq:34} and~\eqref{eq:36}, (b) is derived based on the Bayes’ rule. In~\eqref{eq:78}, $P\left( {\left. V \right|{W_0}} \right)$ and $P\left( {\left. V \right|{W_1}} \right)$  are given by
\vspace{-0.5\baselineskip}
\begin{equation}\label{eq:79}
P\left( {\left. V \right|{W_0}} \right) = \frac{1}{{\sqrt {2\pi \sigma _{\mathit{MCC}}^2} }}\exp \left( {\frac{-{{{V}^2}}}{{2\sigma _{\mathit{MCC}}^2}}} \right)
\end{equation}
\vspace{-1\baselineskip}
\begin{equation}\label{eq:80}
\begin{array}{l}
P\left( {\left. V \right|{W_1}} \right) = P\left( {V\left| {\sum\limits_{j = 1}^M {{X_j}}  \ge G} \right.} \right) = P\left( {V\left| {\sum\limits_{j = 1}^M {{X_j} = G \cup } .... \cup \sum\limits_{j = 1}^M {{X_j} = MG} } \right.} \right)\\
\mathop  = \limits^{({\rm{a}})} \frac{{P\left( {\left. {\sum\limits_{j = 1}^M {{X_j} = G \cup } .... \cup \sum\limits_{j = 1}^M {{X_j} = MG} } \right|V} \right)P\left( V \right)}}{{P\left( {\sum\limits_{j = 1}^M {{X_j} = G \cup } .... \cup \sum\limits_{j = 1}^M {{X_j} = MG} } \right)}}{\rm{ }}\mathop  = \limits^{(b)} \frac{{P\left( {\left. {\sum\limits_{j = 1}^M {{X_j} = G \cup } .... \cup \sum\limits_{j = 1}^M {{X_j} = MG} } \right|V} \right)P\left( V \right)}}{{1 - P(\sum\limits_{j = 1}^M {{X_j}}  = 0)}}\\
 = \frac{{P\left( {\left\{ {\sum\limits_{j = 1}^M {{X_j} = G \cup } .... \cup \sum\limits_{j = 1}^M {{X_j} = MG} } \right\} \cap V} \right)}}{{1 - P(\sum\limits_{j = 1}^M {{X_j}}  = 0)}}\mathop {{\rm{  }} = }\limits^{(c)} \frac{{P\left( {\left( {\left\{ {\sum\limits_{j = 1}^M {{X_j} = G} } \right\} \cap V} \right) \cup .... \cup \left( {\left\{ {\sum\limits_{j = 1}^M {{X_j} = MG} } \right\} \cap V} \right)} \right)}}{{1 - P(\sum\limits_{j = 1}^M {{X_j}}  = 0)}}\\
\mathop {{\rm{  }} = }\limits^{(d)} \frac{{\sum\nolimits_{l = 1}^M {P\left( {\left\{ {\sum\limits_{j = 1}^M {{X_j} = lG} } \right\} \cap V} \right)} }}{{1 - P(\sum\limits_{j = 1}^M {{X_j}}  = 0)}}{\rm{  }} = \frac{{\sum\nolimits_{j = 1}^M {P\left( {V\left| {\sum\limits_{j = 1}^M {{X_j} = lG} } \right.} \right)P(\sum\limits_{j = 1}^M {{X_j}}  = lG)} }}{{1 - P(\sum\limits_{j = 1}^M {{X_j}}  = 0)}}\\
\mathop  = \limits^{(e)} \frac{1}{{\left( {1 - {p_0}} \right)\sqrt {2\pi \sigma _{\mathit{MCC}}^2} }}\sum\limits_{l = 1}^M {{p_l}\exp \left( {\frac{-{{{\left( {V - lG} \right)}^2}}}{{2\sigma _{\mathit{MCC}}^2}}} \right)}.
\end{array}
\end{equation}
Here, (a) follows from the Bayes’ rule, (b) follows from definition of event of ${X_j}$, (c) follows from De Morgan’s law, (d) follows since the events $X_j=lG$, $\forall l$ are mutually exclusive, and (e) follows since the noise of MCC is Gaussian. Hence, replacing~\eqref{eq:79} and~\eqref{eq:80} in~\eqref{eq:78} we have
 \vspace{0.5\baselineskip}
\begin{equation}\label{eq:81}
\begin{array}{l}
\left( {1 - {Q_F}} \right)P({H_0})\frac{1}{{\sqrt {2\pi \sigma _{\mathit{MCC}}^2} }}\exp \left( {\frac{-{{{{V }}^2}}}{{2\sigma _{\mathit{MCC}}^2}}} \right) + {Q_F}P({H_0})\frac{1}{{\left( {1 - {p_0}} \right)\sqrt {2\pi \sigma _{\mathit{MCC}}^2} }}\sum\limits_{l = 1}^M {{p_l}\exp \left( {\frac{-{{{\left({V - lG} \right)}^2}}}{{2\sigma _{\mathit{MCC}}^2}}} \right)} \mathop {\mathop  > \limits^{{H_0}} }\limits_ <  \\
\left( {1 - {Q_D}} \right)P({H_1})\frac{1}{{\sqrt {2\pi \sigma _{\mathit{MCC}}^2} }}\exp \left( {\frac{-{{{ {V}}^2}}}{{2\sigma _{\mathit{MCC}}^2}}} \right) + {Q_D}P({H_1})\frac{1}{{\left( {1 - {p_0}} \right)\sqrt {2\pi \sigma _{\mathit{MCC}}^2} }}\sum\limits_{l = 1}^M {{p_l}\exp \left( {\frac{-{{{\left( {V - lG} \right)}^2}}}{{2\sigma _{\mathit{MCC}}^2}}} \right)},
\end{array}
\end{equation}
 \vspace{-0.5\baselineskip}
which simplifies to
\begin{equation}\label{eq:82}
\begin{array}{l}
\left( {\left( {1 - {Q_F}} \right)P({H_0}) - \left( {1 - {Q_D}} \right)P({H_1})} \right)\exp \left( {\frac{-{{V^2}}}{{2\sigma _{\mathit{MCC}}^2}}} \right) + \\
\left( {{Q_F}P({H_0}) - {Q_D}P({H_1})} \right)\frac{1}{{\left( {1 - {p_0}} \right)}}\sum\limits_{l = 1}^M {{p_l}\exp \left( {\frac{-{{{\left( {V - lG} \right)}^2}}}{{2\sigma _{\mathit{MCC}}^2}}} \right)} \mathop {\mathop  > \limits^{{H_0}} }\limits_ <  0.
\end{array}
\end{equation}
The decision region of DGN is then $V\mathop  < \limits^{{H_0}} {V^{\mathit{THR}}}$, where $V^\mathit{THR}$ is derived numerically from~\eqref{eq:82}.
\subsection{Appendix D. Proof of Theorem 3}
The probability of detection, ${P_D}$, is given by
\vspace{-0.5\baselineskip}
\begin{equation}\label{eq:84}
\begin{array}{l}
{P_D} = P \left( {\left. {{V_{{H_1}}}} \right|{H_1}} \right)\mathop  = \limits^{({\rm{a}})} P \left( {\left. {{V_{{H_1}}}} \right|{H_1},{W_0}} \right)P \left( {\left. {{W_0}} \right|{H_1}} \right) + P \left( {\left. {{V_{{H_1}}}} \right|{H_1},{W_1}} \right)P \left( {\left. {{W_1}} \right|{H_1}} \right)\\
\mathop  = \limits^{(b)} P \left( {\left. {{V_{{H_1}}}} \right|{H_1},{W_0}} \right)\left( {1 - {Q_D}} \right) + P \left( {\left. {{V_{{H_1}}}} \right|{H_1},{W_1}} \right){Q_D}\\
\mathop  = \limits^{(c)} P \left( {\left. {{V_{{H_1}}}} \right|{H_1},\sum\nolimits_{j = 1}^M {{X_j} = 0} } \right)\left( {1 - {Q_D}} \right) + P \left( {\left. {{V_{{H_1}}}} \right|{H_1},\sum\nolimits_{j = 1}^M {{X_j} \ge G} } \right){Q_D}\\
\mathop  = \limits^{(d)} P \left( {\left. {{V_{{H_1}}}} \right|{H_1},\sum\nolimits_{j = 1}^M {{X_j} = 0} } \right)\left( {1 - {Q_D}} \right) + \frac{{\sum\nolimits_{l = 1}^M {P \left( {\left. {{V_{{H_1}}}} \right|\sum\limits_{j = 1}^M {{X_j} = lG,{H_1}} } \right)P \left( {\left. {\sum\limits_{j = 1}^M {{X_j}}  = lG} \right|{H_1}} \right)} }}{{1 - P \left( {\left. {\sum\limits_{j = 1}^M {{X_j} = 0} } \right|{H_1}} \right)}}{Q_D}\\
 = {{\cal Q}}\left( {\frac{{{V^{\mathit{THR}}}}}{{{\sigma _{\mathit{MCC}}}}}} \right)\left( {1 - {Q_D}} \right) + \frac{{\sum\nolimits_{l = 1}^M {{{\cal Q}}\left( {\frac{{{V^{\mathit{THR}}} - lG}}{{{\sigma _{\mathit{MCC}}}}}} \right){{p'}_l}} }}{{1 - {{p'}_0}}}{Q_D},
\end{array}
\end{equation}
where, $V_{H1}$ is the event of $V>V^{\rm{THR}}$, (a) follows from the law of total probability, (b) is derived based on definitions of ${Q_F}$  and ${Q_D}$ respectively in~\eqref{eq:34} and~\eqref{eq:36}, (c) is derived based on definitions of ${W_0}$  and ${W_1}$  in~\eqref{eq:77}, and ${p'_l} = \Pr \left\{ {\left. {U = lG} \right|{H_1}} \right\}$, $l \in \left[ {0,M} \right]$ is given by~\eqref{eq:45}, and (d) is derived as follows,
\begin{equation}\label{eq:85}
\begin{array}{l}
P \left( {{V_{{H_1}}}\left| {{H_1},\sum\limits_{j = 1}^M {{X_j}}  \ge G} \right.} \right) = P\left( {{V_{{H_1}}}\left| {{H_1} \cap \left( {\sum\limits_{j = 1}^M {{X_j} = G \cup } .... \cup \sum\limits_{j = 1}^M {{X_j} = MG} } \right)} \right.} \right)\\
\mathop {\rm{ = }}\limits^{\left( {\rm{a}} \right)} \frac{{P\left( {\left. {\left( {\sum\limits_{j = 1}^M {{X_j} = G \cup } .... \cup \sum\limits_{j = 1}^M {{X_j} = MG} } \right) \cap {H_1}} \right|{V_{{H_1}}}} \right)P\left( {{V_{{H_1}}}} \right)}}{{P\left( {\left( {\sum\limits_{j = 1}^M {{X_j} = G \cup } .... \cup \sum\limits_{j = 1}^M {{X_j} = MG} } \right) \cap {H_1}} \right)}}\mathop {{\rm{  }} = }\limits^{\left( b \right)} \frac{{P\left( {\left( {\left\{ {\sum\limits_{j = 1}^M {{X_j} = G} } \right\} \cap {V_{{H_1}}} \cap {H_1}} \right) \cup .... \cup \left( {\left\{ {\sum\limits_{j = 1}^M {{X_j} = MG} } \right\} \cap {V_{{H_1}}} \cap {H_1}} \right)} \right)}}{{P\left( {\left( {\sum\limits_{j = 1}^M {{X_j} = G \cup } .... \cup \sum\limits_{j = 1}^M {{X_j} = MG} } \right) \cap {H_1}} \right)}}\\
\mathop  = \limits^{\left( c \right)} \frac{{\sum\nolimits_{l = 1}^M {P\left( {\left\{ {\sum\limits_{j = 1}^M {{X_j} = lG} } \right\} \cap {V_{{H_1}}} \cap {H_1}} \right)} }}{{P\left( {\left( {\sum\limits_{j = 1}^M {{X_j} = G \cup } .... \cup \sum\limits_{j = 1}^M {{X_j} = MG} } \right) \cap {H_1}} \right)}}\mathop  = \limits^{\left( d \right)} \frac{{\sum\nolimits_{l = 1}^M {P\left( {{V_{{H_1}}}\left| {\sum\limits_{j = 1}^M {{X_j} = lG \cap {H_1}} } \right.} \right)P\left( {\sum\limits_{j = 1}^M {{X_j}}  = lG \cap {H_1}} \right)} }}{{P\left( {\left( {\sum\limits_{j = 1}^M {{X_j} = G \cup } .... \cup \sum\limits_{j = 1}^M {{X_j} = MG} } \right) \cap {H_1}} \right)}}\\
\mathop  = \limits^{\left( e \right)} \frac{{\sum\nolimits_{l = 1}^M {P\left( {{V_{{H_1}}}\left| {\sum\limits_{j = 1}^M {{X_j} = lG \cap {H_1}} } \right.} \right)P\left( {\left. {\sum\limits_{j = 1}^M {{X_j}}  = lG} \right|{H_1}} \right)P\left( {{H_1}} \right)} }}{{P\left( {\left. {\left( {\sum\limits_{j = 1}^M {{X_j} = G \cup } .... \cup \sum\limits_{j = 1}^M {{X_j} = MG} } \right)} \right|{H_1}} \right)P\left( {{H_1}} \right)}}\mathop  = \limits^{\left( f \right)} \frac{{\sum\nolimits_{l = 1}^M {P\left( {{V_{{H_1}}}\left| {\sum\limits_{j = 1}^M {{X_j} = lG \cap {H_1}} } \right.} \right)P\left( {\left. {\sum\limits_{j = 1}^M {{X_j}}  = lG} \right|{H_1}} \right)} }}{{1 - P\left( {\left. {\sum\limits_{j = 1}^M {{X_j} = 0} } \right|{H_1}} \right)}}.
\end{array}
\end{equation}
\vspace{-0.5\baselineskip}
The steps in deriving~\eqref{eq:85} are similar to those in~\eqref{eq:80}. In a similar way, the NADS probability of false alarm may be computed as
\vspace{-0.5\baselineskip}
\begin{equation}\label{eq:86}
\begin{array}{l}
{P_F} = P\left( {\left. {{V_{{H_0}}}} \right|{H_0}} \right) = P\left( {\left. {{V_{{H_0}}}} \right|{H_0},{W_0}} \right)P\left( {\left. {{W_0}} \right|{H_0}} \right) + P\left( {\left. {{V_{{H_0}}}} \right|{H_0},{W_1}} \right)P\left( {\left. {{W_1}} \right|{H_0}} \right)\\
 = P\left( {\left. {{V_{{H_0}}}} \right|{H_0},{W_0}} \right)\left( {1 - {Q_F}} \right) + P\left( {\left. {{V_{{H_0}}}} \right|{H_0},{W_1}} \right){Q_F}\\
  = P\left( {\left. {{V_{{H_0}}}} \right|{H_0},\sum\nolimits_{j = 1}^M {{X_j} = 0} } \right)\left( {1 - {Q_F}} \right) + P\left( {\left. {{V_{{H_0}}}} \right|{H_0},\sum\nolimits_{j = 1}^M {{X_j} \ge G} } \right){Q_F}\\
P\left( {{V_{{H_0}}}\left| {{H_0},\sum\nolimits_{j = 1}^M {{X_j} = 0} } \right.} \right)\left( {1 - {Q_F}} \right) + \frac{{\sum\nolimits_{l = 1}^M {P\left( {{V_{{H_0}}}\left| {\sum\limits_{j = 1}^M {{X_j} = lG,{H_0}} } \right.} \right)P\left( {\left. {\sum\limits_{j = 1}^M {{X_j}}  = lG} \right|{H_0}} \right)} }}{{1 - P\left( {\left. {\sum\limits_{j = 1}^M {{X_j} = 0} } \right|{H_0}} \right)}}{Q_F}\\
 = {{\cal Q}}\left( {\frac{{{V^{\mathit{THR}}}}}{{{\sigma _{\mathit{MCC}}}}}} \right)\left( {1 - {Q_F}} \right) + \frac{{\sum\nolimits_{l = 1}^M {{{\cal Q}}\left( {\frac{{{V^{\mathit{THR}}} - lG}}{{{\sigma _{\mathit{MCC}}}}}} \right){{p''}_l}} }}{{1 - {{p''}_0}}}{Q_F}.
\end{array}
\end{equation}
where, $V_{H0}$ is the event of $V<V^{\rm{THR}}$, ${p''_l}=\Pr \left\{ {\left. {U = lG} \right|{H_0}} \right\}$, $l \in \left[ {1,M} \right]$ is given by~\eqref{eq:46} and the steps in deriving~\eqref{eq:86} are similar to those in~\eqref{eq:84}.
\subsection{Proof of Lemma 2}
We consider a homogenous molecular environment, if we assume ${\widehat \mathit{NR}_1} \approx {\widehat \mathit{NR}_2} \approx ... \approx {\widehat \mathit{NR}_M} \approx {\widehat \mathit{NR}}$, $p'_l$ in ~\eqref{eq:45} may be approximated as follows
\vspace{-0.5\baselineskip}
\begin{equation}\label{eq:87}
\begin{array}{l}
{{p'}_l} \approx \left( {\begin{array}{*{20}{c}}
M\\
l
\end{array}} \right){\left( {2\pi } \right)^{{{ - M} \mathord{\left/
 {\vphantom {{ - M} 2}} \right.
 \kern-\nulldelimiterspace} 2}}}\sigma _D^{ - M}{\left| {{\Omega ^{\mathit{SC}}}} \right|^{ - {1 \mathord{\left/
 {\vphantom {1 2}} \right.
 \kern-\nulldelimiterspace} 2}}}\int_A {\exp \left( { - \frac{l}{{2M\sigma _D^2}}{{\left( {\widehat \mathit{NR} - \mathit{NR}} \right)}^2}{{\left[ {\bf{1}} \right]}^\dag }{\Omega ^{\mathit{SC}^{- 1}}}\left[ {\bf{1}} \right]} \right)d\widehat \mathit{NR}} \\
{\rm{                                                   }}\int_{{A^C}} {\exp \left( { - \frac{{M - l}}{{2M\sigma _D^2}}{{\left( {\widehat \mathit{NR} - {\mathit{NR}}} \right)}^2}{{\left[ {\bf{1}} \right]}^\dag }{\Omega ^{\mathit{SC}^{-1}}}\left[ {\bf{1}} \right]} \right)d\widehat \mathit{NR}},
\end{array}
\end{equation}
\vspace{-0.5\baselineskip}
where, $\left[ {\bf{1}} \right] = {\left[ {1,...,1} \right]_{1 \times M}}$. Using Holder's inequality~\cite{58} in the RHS of above equation, we have
\begin{equation}\label{eq:88}
\begin{array}{l}
\int_A {\exp \left( { - \frac{l}{{2M\sigma _D^2}}{{\left( {\widehat \mathit{NR} - \mathit{NR}} \right)}^2}{{\left[ {\bf{1}} \right]}^\dag }{\Omega ^{{\mathit{SC}^{ - 1}}}}\left[ {\bf{1}} \right]} \right)d\widehat \mathit{NR}}  \times \\
\int_{{A^C}} {\exp \left( { - \frac{{M - l}}{{2M\sigma _D^2}}{{\left( {\widehat \mathit{NR} - \mathit{NR}} \right)}^2}{{\left[ {\bf{1}} \right]}^\dag }{\Omega ^{{\mathit{SC}^{ - 1}}}}\left[ {\bf{1}} \right]} \right)d\widehat \mathit{NR}}  \le \\
{\left\{ {\int_A {\exp \left( { - \frac{1}{{2\sigma _D^2}}{{\left( {\widehat \mathit{NR} - {\mathit{NR}}} \right)}^2}} \right)d\widehat \mathit{NR}} } \right\}^{\frac{l}{M}{{\left[ {\bf{1}} \right]}^\dag }{\Omega ^{{\mathit{SC}^{ - 1}}}}\left[ {\bf{1}} \right]}} \times \\
{\left\{ {\int_{{A^C}} {\exp \left( { - \frac{1}{{2\sigma _D^2}}{{\left( {\widehat \mathit{NR} - {\mathit{NR}}} \right)}^2}} \right)} d\widehat \mathit{NR}} \right\}^{\frac{{M - l}}{M}{{\left[ {\bf{1}} \right]}^\dag }{\Omega ^{{\mathit{SC}^{ - 1}}}}\left[ {\bf{1}} \right]}}=\\
{\left( {\sqrt {2\pi \sigma _D^2} } \right)^{\left[ {\bf{1}} \right]{\Omega ^{{\mathit{SC}^{ - 1}}}}{{\left[ {\bf{1}} \right]}^T}}}{\left\{ {\int_A {\frac{1}{{\left( {\sqrt {2\pi } {\sigma _D}} \right)}}\exp \left( { - \frac{1}{{2\sigma _D^2}}{{\left( {\widehat \mathit{NR} - {\mathit{NR}}} \right)}^2}} \right)d{\widehat \mathit{NR}}} } \right\}^{\frac{l}{M}{{\left[ {\bf{1}} \right]}^\dag }{\Omega ^{{\mathit{SC}^{ - 1}}}}\left[ {\bf{1}} \right]}} \times \\
{\left\{ {\int_{{A^C}} {\frac{1}{{\left( {\sqrt {2\pi } {\sigma _D}} \right)}}\exp \left( { - \frac{1}{{2\sigma _D^2}}{{\left( {\widehat \mathit{NR} - \mathit{NR}} \right)}^2}} \right)d{\widehat \mathit{NR}}} } \right\}^{\frac{{M - l}}{M}{{\left[ {\bf{1}} \right]}^\dag }{\Omega ^{{\mathit{SC}^{ - 1}}}}\left[ {\bf{1}} \right]}}.
\end{array}
\end{equation}
Considering only the first component in Taylor expansion of the first term in RHS of the above inequality
\vspace{-0.5\baselineskip}
\begin{equation}\label{eq:90}
{p'_l}  \approx  \left( {\begin{array}{*{20}{c}}
M\\
l
\end{array}} \right){\left( {1 - P_D^{NCC}} \right)^{\frac{{M - l}}{M}\left( {\left[ {\bf{1}} \right]^\dag{\Omega ^{S{C^{ - 1}}}}\left[ {\bf{1}} \right]} \right)}}{\left( {P_D^{NCC}} \right)^{\frac{l}{M}\left( {{{\left[ {\bf{1}} \right]}^\dag }{\Omega ^{S{C^{ - 1}}}}\left[ {\bf{1}} \right]} \right)}}.
 \end{equation}
 The RHS of~\eqref{eq:90} is denoted by ${\tilde p'_l}$ and serves as an efficient approximation of ${p'_l}$. The fitting parameter $\alpha$ is obtained numerically for best approximation (See Section V).
In a similar manner ${\tilde p''_l}$  in~\eqref{eq:46} may be calculated. In~\eqref{eq:53} ${\tilde p_l}$ may be calculated simply by replacing ${p'_l}$  and ${p''_l}$ with ${\tilde p'_l}$  and ${\tilde p''_l}$ in~\eqref{eq:42}.
\bibliographystyle{IEEETCOM}
{\footnotesize
\bibliography{IEEENanoBioScience}}

\end{document}